\documentclass[11pt,a4paper]{iopart}
\usepackage[utf8]{inputenc}
\expandafter\let\csname equation*\endcsname\relax

\expandafter\let\csname endequation*\endcsname\relax 
\usepackage{amsmath}
\usepackage{amsfonts}
\usepackage{amssymb}
\usepackage{makeidx}
\usepackage{iopams}
\usepackage{hyperref}

\usepackage{amsthm}
\usepackage{mathrsfs}
\usepackage{subfigure}

\usepackage{todonotes}

\newcommand{\opunit}{\text{1}\kern-0.22em\text{l}}

\usepackage{color}
\definecolor{dgreen}{rgb}{0,0.7,0}

\newcommand{\ie}{\textit{i.e.}}

\def\bea{\begin{eqnarray}}
\def\eea{\end{eqnarray}}
\def\ba{\begin{array}}
\def\ea{\end{array}}
\def\n{\nonumber}
\def\la{\langle}
\def\ra{\rangle}

\usepackage{color,soul}

\begin{document}

\title{Symmetric Exclusion Process  under  Stochastic Power-law Resetting}

\author{Seemant Mishra}
\address{International Centre for Theoretical Sciences, Tata Institute of Fundamental Research, Bengaluru 560089, India}
\address{Indian Institute of Science Education and Research, Pune 411008, India}
\ead{seemant.mishra@icts.res.in}

\author{Urna Basu}
\address{S. N. Bose National Centre for Basic Sciences, Kolkata 700106, India}
\ead{urna@bose.res.in}

\begin{abstract}

We study the behaviour of a symmetric exclusion process in the presence of non-Markovian stochastic resetting, where the configuration of the system is reset to a step-like profile at power-law waiting times with an exponent $\alpha$. We find that the power-law resetting leads to a rich behaviour for the currents, as well as density profile. We show that, for any finite system, for $\alpha<1$, the density profile eventually becomes uniform while for $\alpha >1$, an eventual non-trivial stationary profile is reached. We also find that, in the limit of thermodynamic system size, at late times, the average diffusive current grows $\sim t^\theta$ with $\theta = 1/2$ for $\alpha \le 1/2$, $\theta = \alpha$ for $1/2 < \alpha \le 1$ and $\theta=1$ for $\alpha > 1$. We also analytically characterize the distribution of the diffusive current in the short-time regime using a trajectory-based perturbative approach. Using numerical simulations, we show that in the long-time regime, the diffusive current distribution follows a scaling form with an $\alpha-$dependent scaling function. We also characterise the behaviour of the total current using renewal approach.  We find that the average total current also grows algebraically $\sim t^{\phi}$ where $\phi = 1/2$ for $\alpha \le 1$, $\phi=3/2-\alpha$ for $1 < \alpha \le 3/2$, while for $\alpha > 3/2$ the average total current reaches a stationary value, which we compute exactly. The variance of the total current also shows an algebraic growth with an exponent $\Delta=1$ for $\alpha \le 1$, and $\Delta=2-\alpha$ for $1 < \alpha \le 2$, whereas it approaches a constant value for $\alpha>2$. The total current distribution remains non-stationary for $\alpha<1$, while, for $\alpha>1$, it reaches a non-trivial and strongly non-Gaussian stationary distribution, which we also compute using the renewal approach.

\end{abstract}

\maketitle

\section{Introduction}

Stochastic Resetting refers to the process of intermittently stopping and restarting the dynamical evolution of a system \cite{SRApplications}. The study of stochastic resetting has gained immense interest recently, owing to its potential applications in a wide range of fields ranging from search problems \cite{search1,search4}, population dynamics \cite{population1,population2} to computer science~\cite{comp_restarts,search2} and financial markets~\cite{stock,stock2, taxresetting}. Perhaps the simplest example of resetting dynamics is when the position of a Brownian particle is reset to a fixed point with a constant rate~\cite{Brownian}. Despite the apparent simplicity of the dynamics, it leads to a set of intriguing behaviour which includes the emergence of a non-trivial stationary state, a dynamical transition in the relaxation process as well as a finite mean first-passage time~\cite{relaxation}. Various generalizations and extensions of this simple model have been studied over the past decade \cite{Brownian3, absorption,highd,Mendez2016,Puigdellosas,deepak,Arnab2019_2}.  Examples include resetting of a diffusing particle in a potential \cite{reset_potential,potential2}, in the presence of an absorbing target whose position is drawn from a distribution \cite{diff_optimal_reset}, or in the presence of clusters of targets \cite{DiffCluster}. The role of different resetting protocols has also been explored in the literature, for example the effects of resets with a refractory period \cite{refractory}, and non-instantaneous resets \cite{noninstreset, external_pot_reset}. Certain protocols have also been utilised to experimentally realise stochastic resetting \cite{experimental_reset}.

A particularly interesting example is that of non-Markovian resetting protocols, where the waiting times between consecutive resetting events do not follow an exponential distribution. The effect of such non-Markovian resetting with power-law waiting times on Brownian motion has been studied recently, which shows that there exists a regime of parameters for which the position distribution of the particle does not approach a stationary state~\cite{Diff_plr}. A similar resetting protocol is also shown to lead to a much richer dynamical behaviour for extended systems --- a fluctuating surface subjected to power-law resetting can show bounded or unbounded growth of the interface width, depending on the choice of parameters \cite{Gupta2016}. Another example is totally asymmetric exclusion process under power-law resetting, which can give rise to non-monotonic and non-stationary density profiles in the long-time limit~\cite{TASEPwsr}. However, the effect of such non-Markovian resetting protocols on the transport properties of extended systems has not been studied so far.

In this paper, we study the effect of power-law resetting on the transport properties of symmetric simple exclusion process. It has been shown recently that, the symmetric exclusion process dynamics under Markovian resetting to a step-like configuration leads to a drastic change in the behaviour of the particle current flowing through the system  \cite{SEPwsr,SEPdicho}. We show that the power-law resetting leads to a richer behaviour of the current and density profiles. In particular, we show that for any finite $L$, when the power-law exponent $\alpha<1$, the density profile does not approach an eventual non-trivial stationary state and relaxes to the uniform profile. For $\alpha>1$, however, we find that the density approaches a non-trivial stationary profile. We also show that the in the limit of thermodynamically large system size, the average diffusive current grows algebraically with time with an exponent $\theta(\alpha)$, which we compute exactly. We further develop a perturbative approach to compute the diffusive current distribution at short times.  For longer times, we find that, the diffusive current distributions have a scaling form, with an $\alpha$--dependent scaling function which approaches Gaussian for large $\alpha$. This is in contrast with SEP without resetting as well as SEP in the presence of Markovian resetting --- the typical fluctuations of the diffusive current are always Gaussian in these scenarios.  We also find that the average total current, which, similar to the diffusive current, also shows an algebraic growth in time with a different exponent $\phi(\alpha)$, which we compute exactly. For $\alpha<1$, the total current does not approach a stationary distribution, and shows strongly non-Gaussian fluctuations. For $\alpha>1$, however, the distributions approach a stationary distribution, which we also compute using the renewal approach.

\section{Model and Results}\label{sec:model}

The Symmetric Exclusion Process (SEP) is a simple model for the transport of particles in one dimension \cite{Spitzer,Liggett}. This model has been used to describe various physical phenomena including the motion of molecular motors, the movement of ions in a porous medium, and transport in narrow channels. In this section, we define the rules of the evolution of SEP under a power law resetting, and present a summary of our main results.

Let us consider a one-dimensional lattice of size $L$, with periodic boundary conditions. Each site $x$ of the lattice can contain at most one particle. The particles can hop to their left or right nearest neighbours with a constant rate (which can be taken to be $1$ without loss of any generality). We can denote the occupancy of a site $x$ by a variable $s_x$, which takes the values $1$ if the site is occupied and $0$ if the site is not occupied. Then the configuration of the system is symbolically represented by ${\cal C} = \{s_x; x=0,1,2,3.....,L-1\}$.  

In the absence of resetting,  the configuration of SEP evolves via symmetric hopping of particles to vacant nearest neighbour sites. In the presence of a power-law resetting, the configuration of the system evolves via two different kinds of dynamical moves:
\begin{itemize}
\item {\bf Hopping}: A randomly chosen particle can hop to one of its nearest neighbouring sites with a unit rate, provided that the target site is empty.

\item {\bf Resetting}: The system is intermittently reset to a fixed configuration ${\cal C}_0$.  In the following, we consider ${\cal C}_0$ to be a step-like configuration where all the particles are in the left half of the system, \ie,
\bea
{{\cal C}_0}:=\left \{ \begin{aligned}
                   s_x=1~& ~\text{for}~0 \le x \le \frac L2-1,\\
                   s_x=0 ~&~\text{otherwise.}  
                    \end{aligned}
 \right.
 \label{eq:C0}
\eea
\end{itemize}
Furthermore, we consider the case where the initial configuration of the system is also ${\cal C}_0$. The resetting protocol we study is a specific non-Markovian one where the waiting time $\tau$ between two consecutive resetting events is drawn from a power-law distribution,
\begin{align}
    \label{eq:pld}
\psi(\tau) =  \frac{\alpha}{t_0} \left(\frac{\tau}{t_0}\right)^{-(1+\alpha)} \qquad \alpha>0, \: t\in[t_0,\infty),
\end{align}
where $t_0 >0$. It is to be noted that all moments of $\psi(\tau)$ are infinite for $\alpha <1$. For $\alpha >1$ the mean waiting-time $\la \tau \ra = t_0 \alpha /(\alpha -1)$ becomes finite  while for $\alpha >2$ the second moment $\la \tau^2 \ra =t_0^2 \alpha /(\alpha -2)$  also becomes finite. 

The divergence of the mean and variance of $\tau$ in certain parameter regimes makes the power-law resetting protocol drastically different than the constant rate resetting, which corresponds to an exponential waiting-time distribution, with a finite mean and variance. Even a single diffusing particle under power-law resetting shows a spectrum of rich long-time behaviour including a non-diffusive spreading for $\alpha <1$ and a nontrivial stationary state for $\alpha >1$ ~\cite{Diff_plr}.

In the absence of resetting, the time evolution of SEP starting from the step-like configuration ${\cal C}_0$ has been studied in Ref.~\cite{SEPstep} where it was shown that the average time-integrated diffusive particle current across the central bond increases as $\sim \sqrt t$ with time $t$. The behaviour of SEP, in the presence of Markovian resetting, has been investigated in \cite{SEPwsr} where the configuration resets to ${\cal C}_0$ at a constant rate. In this case, the average diffusive current was found to grow much faster, namely, $\sim t$. Moreover, the presence of the resetting gives rise to another current, namely the total current, which accounts for the net flow of particles across the central bond arising from both the hopping and resetting dynamics. The total current approaches a stationary distribution with strong non-Gaussian fluctuations.

\begin{figure}
\centering \includegraphics[width=14cm]{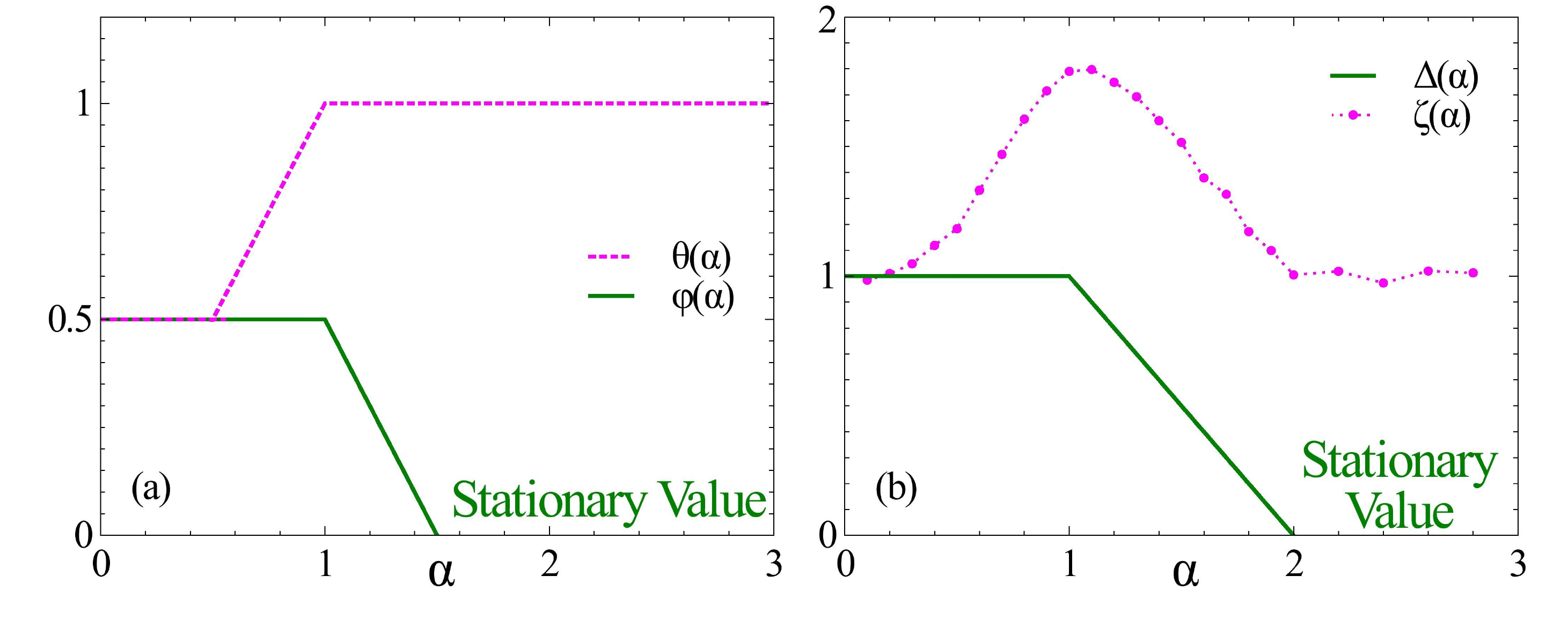} 
\caption{Plot of the exponents (a) $\theta(\alpha)$ and $\phi(\alpha)$ characterizing the late-time growth of the average, and (b) $\zeta(\alpha)$ and $\Delta(\alpha)$ characterizing the growth of the variance, of diffusive and total currents respectively, as a function of the power-law exponent $\alpha$.}\label{fig:exponents_moments} 
\end{figure}

In this work, we explore SEP under non-Markovian stochastic resetting, with the waiting times drawn from a power-law distribution with an exponent $\alpha$. We focus on the behaviour of the density profile, diffusive current, and total current. Before going into the details of the computations,  a brief summary of the main results is presented below. 

\begin{itemize}
\item {{\bf Density Profile:}} First, we consider the time-evolution of the average density profile $\rho(x,t) = \la s_x(t) \ra$. Using a renewal approach, we show that for $\alpha > 1$, the density eventually reaches an inhomogeneous stationary profile, which we compute exactly \eqref{eq: stat_density_profile_ag1}. For $\alpha \le 1$, for any finite $L$, the stationary density profile becomes uniform with $\rho(x)=1/2$ at all sites.

\item {\bf Diffusive current:} The non-Markovian resetting drastically affects the behaviour of the particle current flowing through the system. We first focus on the diffusive current $J_\text{d}(t)$, which measures the number of particles crossing the central bond during the time interval $[0,t]$ due to the hopping dynamics. Using the analytical expression for the time-dependent density profile, we compute the average diffusive current $\la J_\text{d}(t) \ra$. We find that, in the limit of thermodynamically large system size $L \to \infty$, and at late-times ($t \gg t_0 $), the diffusive current grows algebraically,  with an $\alpha$ dependent exponent (see Fig.~\ref{fig:exponents_moments}) 
    \bea
  \la J_\text{d}(t) \ra \sim t^{\theta(\alpha)} \quad \text{with}~~ \theta(\alpha)  = \left \{ \begin{split}
  \frac 12 & ~~ \text{for} ~ \alpha \le \frac 12 \cr
  \alpha & ~~ \text{for} ~ \frac 12 < \alpha \le 1 \cr
  1 & ~~ \text{for} ~ \alpha > 1 \cr
  \end{split}
  \right. \label{eq:Jdav_t}
    \eea

To characterise the fluctuations of $J_\text{d}(t)$, we measure its variance  
$\sigma^2_\text{d}(t)= \la J_\text{d}(t)^2 \ra - \la J_\text{d}(t) \ra^2$ using numerical simulations. We find that,  at late times,  the variance also shows an algebraic growth 
$\sigma^2_\text{d}(t) \sim t^{\zeta(\alpha)}$. The exponent $\zeta(\alpha)$  is estimated numerically, as shown in Fig. \ref{fig:exponents_moments}.

\item The unusual nature of the fluctuations of the diffusive current becomes clearer from the behaviour of the distribution $P(J_\text{d},t)$. In the short-time regime ($ t \sim t_0$) we compute $P(J_\text{d},t)$ analytically, using a trajectory-based perturbative approach, which shows strongly non-Gaussian features. Using extensive numerical simulations, we also show that, in the long-time regime, the diffusive current distribution admits a scaling form,
\bea
P(J_\text{d},t ) = \frac 1{\sigma_\text{d}(t)} {\cal Q}_\alpha\left( \frac{J_\text{d} - \la J_\text{d}(t) \ra  }{\sigma_\text{d}(t)} \right)
\eea
where the shape of the scaling function ${\cal Q}_\alpha(z)$ strongly depends on $\alpha$ for $\alpha<2$, whereas for $\alpha >2$, the scaling function approaches a Gaussian in the large-time regime.

\item {\bf Total Current:} Apart from the diffusing current, the particle motion due to the resetting gives rise to an additional current \cite{SEPwsr}.  We study the behaviour of the total current $J(t)$, which measures the net transport of particles (through hopping and resetting) across the central bond during the interval $[0,t]$.  We find that, in the thermodynamic limit, the average total current $\la J(t) \ra$ eventually approaches a stationary value for $\alpha > 3/2$, which we compute exactly \eqref{eq:Jtstat}. On the other hand, for $\alpha \le 3/2$,  $\la J(t) \ra$ shows an algebraic growth at late-times,
\bea
  \la J(t) \ra \sim t^{\phi(\alpha)} \quad \text{with}~~ \phi(\alpha)  = \left \{ \begin{split}
  \frac 12 \quad & ~~ \text{for} ~ \alpha \le  1 \cr
  \frac 32 - \alpha & ~~ \text{for} ~ 1 < \alpha \le \frac 32 
  \end{split}
  \right. \label{eq:Jav_growth}
\eea

\item We also characterise the fluctuations of the total current through the renewal approach. We find that the variance of the total current $\sigma^2(t)$ too grows algebraically, 
\bea
\sigma^2(t) \sim t^{\Delta(\alpha)} \quad \text{with}~~ \Delta(\alpha)  = \left \{ \begin{split}
  1 \quad & ~~ \text{for} ~ \alpha \le  1 \cr
  2-\alpha & ~~ \text{for} ~ 1 < \alpha < 2 
  \end{split}
  \right.
\eea

\item For $\alpha<1$, we obtain the time-dependent distributions using the renewal equation approach, which can be evaluated numerically. The distributions are strongly non-Gaussian and show some interesting behaviour near $J=0$. For $\alpha>1$ the total current approaches a stationary distribution which shows typical non-Gaussian fluctuations.

\end{itemize}

The rest of the paper is organised as follows. In Section \ref{sec:RnTb} we briefly introduce the renewal equation and the trajectory-based approaches to non-Markovian resetting. In Sections \ref{sec:density profile},\ref{sec:Diffusive current} and \ref{sec:Total current} we discuss the behaviours of the density profile, the diffusive current, and the total current, respectively. We conclude with some open questions in Section ~\ref{sec:conclusion}.

\section{Non-markovian resetting: Renewal equation and trajectory based approach} \label{sec:RnTb}

The general approach towards understanding systems under stochastic resetting is to consider resets as a renewal process, with the waiting times drawn from a distribution $\psi(\tau)$. The renewal equation corresponding to such a process has been derived in \cite{SRApplications}. In the following, we re-derive the renewal equation in the context of resetting of SEP using a trajectory-based approach. 

Let us consider a trajectory of the system with $n$ resetting events during the interval $[0,t]$. Let $\tau_i$ denote the interval between $(i-1)$-th and $i$-th resetting events and $\tau_{n+1}$ denote the interval between the last resetting event and the final time $t$; clearly, $\sum_{i=1}^{n+1} \tau_i =t$. The probability density for such a trajectory is given by,
\begin{align}
\mathscr{P}_n(\{\tau_i\};t) = \prod_{i=1}^{n}\psi(\tau_i) f(\tau_{n+1})\delta\left(t - \sum_{i=1}^{n+1} \tau_i \right), 
\end{align} 
where the survival function $f(\tau_{n+1}) = \int_{\tau_{n+1}}^{\infty} d\tau~\psi(\tau)$ denotes the probability that no resetting has occurred during the interval $\tau_{n+1}$. The probability that, starting from $ {\cal C}_0$ at $t=0$, the system is in the configuration ${\cal C}$ at time $t$ can be obtained by considering the contributions from all possible trajectories,
\begin{align}
{\cal P}({{\cal C}}, t|{\cal C}_0) = f(t) {\cal P}_0({{\cal C}},t|{\cal C}_0)
+\sum_{n=1}^\infty \int \prod_{i=1}^{n} d \tau_i\,d \tau_{n+1} ~\psi(\tau_i) f(\tau_{n+1}) {\cal P}_0({{\cal C}},\tau_{n+1}|{\cal C}_0) \delta (t-\sum_{i=1}^{n+1} \tau_i).  \label{eq:Pnct}
\end{align}
Here ${\cal P}_0({{\cal C}},t|{\cal C}_0)$ denotes the probability that, {\it in the absence of resetting}, the system reaches the configuration ${\cal C}$ at time $t$, starting from the configuration ${\cal C}_0$. The first term on the right-hand side of \eqref{eq:Pnct} denotes the contribution from the trajectories which do not undergo resetting, while the second term denotes the contributions from the trajectories with at least one resetting.

To proceed further,  it is convenient to take the Laplace transform of \eqref{eq:Pnct} with respect to time $t$, 
\begin{align}
\tilde {\cal P}({{\cal C}}, s|{\cal C}_0) &= \int_0^\infty dt~e^{-s t} {\cal P}({{\cal C}}, t|{\cal C}_0)\cr
&= \int_0^\infty dt~e^{-s t}  f(t) {\cal P}_0({{\cal C}},t|{\cal C}_0)+  
\sum_{n=1}^\infty \tilde \psi(s)^n \int_0^\infty d \tau ~e^{-s \tau}  f(\tau) {\cal P}_0({{\cal C}},\tau|{\cal C}_0) \cr
&=\left (1 + \frac {\tilde \psi(s)}{1 - \tilde \psi(s)}\right) \int_0^\infty d \tau ~e^{-s \tau}  f(\tau) {\cal P}_0({{\cal C}},\tau|{\cal C}_0) \label{eq:Pcs}
\end{align}
where $\tilde \psi(s) = \int_0^\infty d\tau~ e^{-s \tau}\psi(\tau)$ denotes the Laplace transform of the waiting-time distribution. Note that, for later convenience we have kept the contributions from $n=0$ and $n>0$ terms separate.

The last renewal equation for arbitrary resetting protocols is obtained by taking the inverse Laplace transform of \eqref{eq:Pcs},
\bea
{\cal P}({{\cal C}}, t|{\cal C}_0) = f(t) {\cal P}_0({{\cal C}},t|{\cal C}_0) + \int_0^t d\tau~ \gamma(t-\tau) f(\tau) {\cal P}_0({\cal C},\tau| {\cal C}_0)
\label{eq:nonm_renewal_full}
\eea
where $\gamma(t)$ is the inverse Laplace transform of $\frac {\tilde \psi(s)}{1 - \tilde \psi(s)}$ with respect to $s$.  Physically $\gamma(t-\tau)$ denotes the probability that the last reset (irrespective of previous ones) has occurred at time $t-\tau$, although it is often hard to write an explicit expression for it~\cite{SRApplications}.

It is particularly interesting to see whether the non-Markovian resetting eventually leads to a stationary state where configuration probabilities do not evolve with time anymore. In fact, it has been shown that a stationary state exists only when $\int_0^\infty f(\tau) d \tau = \la \tau \ra$ is finite, and the stationary state weight ${\cal P}_{st}({\cal C})$ is most conveniently obtained from Eq.~\eqref{eq:Pcs}  by computing the coefficient of $1/s$ in the  $s \to 0$ limit~\cite{SRApplications}, 
\bea 
{\cal P}_{st}({\cal C}) =  \frac 1 {\la \tau \ra} \int_0^\infty d\tau f(\tau){\cal P}_0({\cal C},\tau|{\cal C}_0). \label{eq: stationary_state}
\eea

For the power-law waiting time distribution given by Eq.~\eqref{eq:pld}, we have the survival function,
\bea    \label{eq:f}
f(\tau) =\int_{\tau}^\infty d \tau' \psi(\tau') =\left \{ \begin{split}
1 & ~~ \text{for}~~ \tau \le t_0 \cr
\left(\frac {\tau}{t_0} \right)^{-\alpha} & ~~\text{for} ~~ \tau > t_0. 
 \end{split}
 \right.
\eea
In this case, $\la \tau \ra$ converges for $\alpha >1$ only. Consequently, we expect that the SEP under power-law resetting will reach a stationary state only for $\alpha >1$.  On the other hand, for $\alpha <1$, the system remains in a transient state. To understand  the behaviour of the system for $\alpha<1$, it is convenient to recast the renewal equation \eqref{eq:nonm_renewal_full} as,
\bea \label{eq: nonm_renewal_F}
\fl {\cal P}({\cal C}, t) =  \int_0^t d \tau F(\tau, t-\tau){\cal P}_0({\cal C},\tau),
\textrm{where} \quad F(\tau, t-\tau) = [\delta(t-\tau) + \gamma(t-\tau)] f(\tau).
\eea
For $\alpha<1$,  an explicit form for $F(\tau, t-\tau)$ in the limit $t \gg t_0$ has been calculated previously~\cite{Diff_plr, FormF},
\bea
\label{eq: F_al1}
F(\tau, t-\tau) = \frac{\sin{(\pi\alpha)}}{\pi \tau}\left (\frac{t}{\tau}-1 \right)^{\alpha-1}.
\eea

In the following sections, we use Eq.~\eqref{eq: nonm_renewal_F} to analyse the transient time-dependent behaviour of the system in the regime $\alpha<1$ and Eq.~\eqref{eq: stationary_state} for the behaviour of observables in the stationary state for $\alpha>1$. To get the time-dependent behaviour of certain quantities, we found that it was easier to work in the Laplace domain, hence in those cases, we use the trajectory-based approach of Eq. \eqref{eq:Pcs}.

\section{Density profile}\label{sec:density profile}

The density profile $\rho(x,t)$ measures the average number of particles at site $x$ at time $t$, 
\begin{align}
 \rho(x,t)= \la s_x(t) \ra = \sum_{\cal C} s_x {\cal P}({\cal C},t).
\end{align} 
From the above equation, it is clear that the density profile must also satisfy the  renewal equation  \eqref{eq:nonm_renewal_full},
\bea \label{eq: nonm_renewal_density}
 \rho(x,t) = f(t) \rho_0(x,t) +  \int_0^t d \tau ~\gamma(t-\tau) f(\tau)  \rho_0(x,\tau).
\eea
Here $\rho_0(x,t)$ is the time-dependent density profile in the absence of resetting, starting from the step-like profile corresponding to ${\cal C}_0$. The explicit form of $\rho_0(x,t)$  can be calculated in a straightforward manner  and is given by~\cite{SEPwsr},
\bea \label{eq: density_profile}
\rho_0(x,t) = \frac{1}{2} + \frac{1}{L}\sum_{n=1,3,5...}^{L-1} \: e^{-i \frac{2 \pi n x}{L}}\left(1 + i \cot{\frac{\pi n}{L}}\right) e^{-\lambda_n t}, 
\eea
with $\lambda_n = 2(1-\cos{\frac{2\pi n}{L}})$. 

In the following, we consider the large time behaviour of the density profile following Eq. \eqref{eq: density_profile} for the two cases $\alpha <1$ and $\alpha >1$ separately.

\subsection{Density profile for $\alpha<1$}

\begin{figure}[t] 
    \centering
    \includegraphics[width=14 cm]{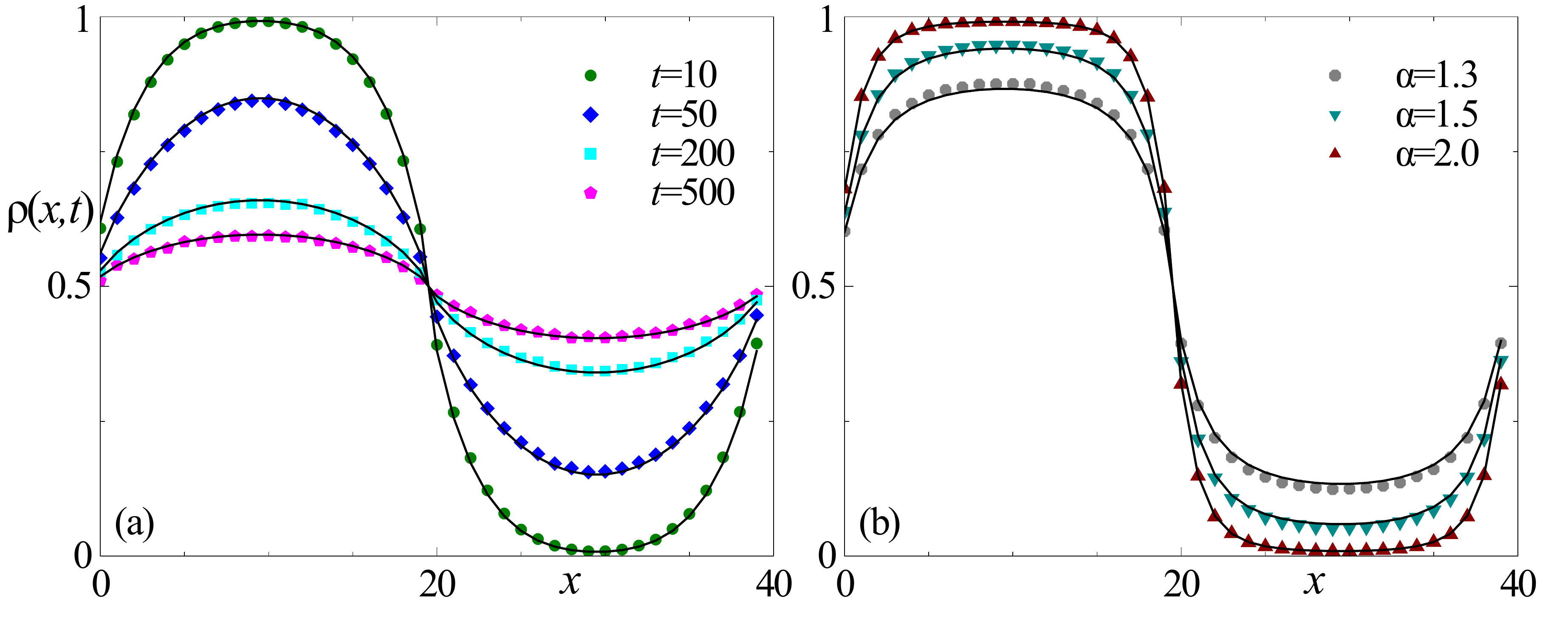}
    \caption{Density profile for SEP under power-law resetting: (a) shows the time evolution of $\rho(x,t)$ for $\alpha=0.5$ and $t_0=0.1$, and (b) shows the stationary profile $\rho(x)$ for a set of different values of $\alpha >1$ and $t_0=1.$ The solid lines indicate the analytical predictions from Eqs.~\eqref{eq: density_profile_al1} and \eqref{eq: stat_density_profile_ag1}, and the symbols indicate the data obtained from numerical simulations performed on a lattice of size $L=40$.}
    \label{fig: denisty_profiles}
\end{figure}

For $\alpha<1$, we use  renewal equation Eq.~\eqref{eq: nonm_renewal_F} for the density profile which leads to
\bea    \label{eq: density_profile_al1}
\rho(x,t) &=& \int_0^t ~d\tau F(\tau,t-\tau)\rho_0(x,\tau) \\ \n
&=& \frac{1}{2} + \frac{1}{L}\sum_{n=1,3,5...}^{L-1} \: e^{-i \frac{2 \pi n x}{L}}(1 + i \cot{\frac{\pi n}{L}}) L_{\alpha-1}(-\lambda_n t).
\eea
where $L_n(z)$ is the Laguerre polynomial (see Sec.~18.3  of \cite{DLMF}). At large times $L_{\alpha-1}(-\lambda_nt)$ decays to zero for all values of $n$. So the only contribution to the density profile comes from the first term. Hence, for all finite values of $L$ the density approaches the trivial uniform profile $\rho(x)=\frac 12$. This result is compared with numerical simulations in Fig.~\ref{fig: denisty_profiles}(a), which shows an excellent agreement.

\subsection{Density profile $\rho(x,t)$ for $\alpha>1$}

For $\alpha>1$, $\rho(x,t)$ approaches a non-trivial stationary profile, which can be exactly evaluated using  Eq. \eqref{eq: stationary_state}, and is given by 
\bea    \label{eq: stat_density_profile_ag1}
\rho(x) &=& \frac{\int_0^\infty~d\tau ~f(\tau) \rho_0(x,\tau)}{\int_0^\infty~d\tau ~f(\tau)} \\ \n
&=& \frac{1}{2} + \frac{1}{L}\sum_{n=1,3,}^{L-1} e^{- 2 \pi i \frac{n x}{L}}\left(1 + i \cot{\frac{\pi n}{L}}\right)\, G(\lambda_n),\n
\eea
with
\bea    \label{eq: ag1_density_profile_G}
G(\lambda_n) = \frac{\alpha-1}{\alpha t_0} \left( \frac{1-e^{-\lambda_n t_0}}{\lambda_n} + t_0 E_\alpha(\lambda_n t_0) \right).
\eea
Here $E_{\nu}(z)$ is the exponential integral function (see Sec.~18.19  of \cite{DLMF}). Figure~\ref{fig: denisty_profiles}(b) shows a comparison of the theoretical predictions from Eq.~\eqref{eq: stat_density_profile_ag1} and data from numerical simulations. A perfect match between these curves confirms our prediction of the stationary density profile.

Qualitatively, the stationary profile for the $\alpha >1$ case is similar to the Markovian resetting scenario, however, quantitatively these are different. Moreover, for the power-law resetting with $\alpha<1$ the stationary profile becomes uniform, which is absent for the Markovian resetting. 

\section{Diffusive current}\label{sec:Diffusive current}

The diffusive current refers to the flow of particles due to the hopping dynamics. We focus on the time-integrated diffusive current across the central bond  $J_\text{d}(t)$, which denotes the net number of particles crossing the central bond $(\frac L2-1, \frac L2)$ from left to right due to hopping, during the time-interval $[0,t]$. Let us denote the corresponding diffusing current for SEP in absence of resetting by $J_0(t)$. In the limit of large system size, and $t \gg 1$, the typical fluctuations of $J_0(t)$ are known to be  Gaussian with mean and variance growing algebraically in time~\cite{SEPstep},
\begin{align}   
P_0(J_0,t) \simeq \frac 1{2\pi \sigma_0^2(t)} \exp{\left[-\frac{(J_0 - \la J_0(t) \ra)^2}{2 \sigma_0^2(t)}\right]} \label{eq:P0}
\end{align}
where, 
\begin{align}   
\la J_0(t) \ra & =  e^{-2t} t \Big[I_0(2t) + I_1(2t) \Big] \simeq  \sqrt{\frac{t}{\pi}}, \quad \text{for} ~~ t \gg 1,\cr
\quad  \text{and,}
\quad \sigma_0^2(t) &\equiv \la J_0^2(t) \ra - \la J_0(t) \ra^2 \simeq \left( 1-\frac{1}{\sqrt{2}} \right)\sqrt{\frac{t}{\pi}}. \label{eq:mean_n_var} 
\end{align}
In the presence of Markovian resetting the average diffusive current shows a linear growth with time~\cite{SEPwsr, SEPdicho}. As we show below, the non-Markovian power-law resetting protocol leads to a more complex and rich behaviour of the diffusive current.

\subsection{Average diffusive current}

First, we focus on the average diffusive current flowing across the central bond. To this end, it is convenient to look at the time-integrated current from a different perspective. The time-integrated current $J_d$ can be expressed as an integral of the instantaneous current,
\bea
J_\text{d}(t) = \int_0^t ~ds ~j_\text{d}(s), \label{eq:Jd_jd}
\eea
where $j_\text{d}(s) ds$ denotes the number of particles crossing the central bond during the interval $[s, s+ds]$. The average instantaneous current is related to the density gradient across the central bond,
\bea \label{eq: avg_j}
\la j_d(t) \ra &= & \left\la s_{\frac L2-1}(1-s_{\frac L2})\right\ra - \left\la (1- s_{\frac L2-1})s_{\frac L2}\right\ra \cr
&= &\rho\left(\frac L2 -1,t\right) - \rho\left(\frac L2,t \right). 
\eea
Clearly, the behaviour of average instantaneous current and, in turn, the average time-integrated diffusive current $\la J_\text{d}(t) \ra$ would be very different for  $\alpha \le 1$ and $\alpha > 1$, owing to the different behaviours of $\rho(x,t)$ in these two regimes.

\subsubsection{Average diffusive current for $\alpha<1$:}

For $\alpha <1$, the average instantaneous diffusive current in the limit $ t \gg t_0$ can be computed using \eqref{eq: avg_j}, along with Eq. \eqref{eq: density_profile_al1}, 
\bea
\la j_d(t) \ra = \frac{2}{L}\sum_{n=1,3,5...}^{L-1} L_{\alpha-1}(-\lambda_n t).
\eea
The average time-integrated diffusive current, in turn, is given by,
\bea \label{eq: diffusive_current_al1}
\la J_\text{d}(t) \ra = \int_0^t ~ds ~\la j_\text{d}(s) \ra=\frac{2}{L}\sum_{n=1,3,5...}^{L-1} \frac{1}{\lambda_n}[L_{\alpha}(-\lambda_n t) - L_{\alpha-1}(-\lambda_n t)].
\eea

For any finite system size $L$, the long-time behaviour of the diffusive current can be obtained by looking at the asymptotic behaviour of $L_\beta(z)$~\cite{DLMF}. For large $z$ and $\beta > -1$,
\bea
L_{\beta}(-z) = \frac{z^{\beta}}{\Gamma(1+\beta)} + O(z^{\beta -1}). 
\eea
Substituting the above in Eq. \eqref{eq: diffusive_current_al1}, we get  the leading order behaviour of the diffusive current in the large time regime, 
\bea    \label{eq: Jd_sum}
\la J_\text{d}(t) \ra \simeq \frac 2L \sum_{n=1,3,5...}^{L-1} \lambda_n^{\alpha -1} \frac {t ^\alpha}{\Gamma(1+\alpha)}.
\eea
For all finite values of $L$, the sum in the above equation converges and hence, the average current grows as $\sim t^\alpha$ at large times for all $\alpha < 1$.
In the limit of thermodynamically large system $L \to \infty$, the sum over $n$ in Eq.~\eqref{eq: Jd_sum} can be converted to an integral over $q=2\pi n/L$, yielding, 
\bea    \label{eq:Jd_thermodynamic_q}
\la J_\text{d}(t) \ra \simeq \int_0^{2\pi} \frac{dq}{2\pi} ~\lambda_q^{\alpha-1}\frac{t^\alpha}{\Gamma(1+\alpha)}, 
\eea
where $\lambda_q= 2(1- \cos q)$. 
This integral converges for $\alpha > \frac 12$, leading to the 
same algebraic large-time growth of the average current as finite systems,
\bea    \label{eq:aghalf}
\la J_\text{d}(t) \ra \simeq \frac{2^{2\alpha-1}\sqrt{\pi}~\Gamma(\alpha-\frac{1}{2})}{\Gamma(\alpha)\Gamma(1+\alpha)}t^\alpha, \qquad \text{for} \quad \alpha > \frac 12. \label{eq:Jd_ag0.5}
\eea

For $\alpha < \frac 12$, however, the integral in Eq.~\eqref{eq:Jd_thermodynamic_q} does not converge and we need to use a different approach. In this case, we go back to the ordinary SEP dynamics and take the $L \to \infty$ limit before using the renewal Eq. \eqref{eq: density_profile_al1}. For ordinary SEP on a thermodynamically large system, the average instantaneous current across the central bond is given by \cite{SEPwsr}, 
\bea
\la j_0(t) \ra &=& e^{-2t}I_0(2t).
\eea
Following  Eq. \eqref{eq: density_profile_al1}, we then have, for $t \gg t_0$,
\bea
\fl \la j_\text{d}(t) \ra = \int_0^t d\tau F(\tau, t -\tau) \la j_0(\tau) \ra = \int_0^t d \tau \frac{\sin (\pi \alpha)}{\pi \tau} \left(\frac t \tau -1 \right)^{\alpha-1} e^{-2\tau}I_0(2 \tau). \label{eq:jd_al1}
\eea

For large t the dominant contribution to the above integral comes from the large $\tau \gg 1$  regime, which leads to,
\bea
\la j_\text{d}(t) \ra \simeq \frac{2}{\pi^2 \sqrt{t}} \, \Gamma\left(\frac 12-\alpha\right) \Gamma(\alpha) \sin(\pi \alpha).\label{eq:insJ_thermodynamic}
\eea
Using the above equation in \eqref{eq:Jd_jd} we get, for $\alpha < \frac 12$,
\bea    \label{eq:alhalf}
\la J_\text{d}(t) \ra \simeq  \frac{4 \sqrt{t}}{\pi^2} \, \Gamma\left(\frac 12-\alpha\right) \Gamma(\alpha) \sin(\pi \alpha). 
\eea
Note that, the integral in Eq.~\eqref{eq:jd_al1} converges only in the regime $\alpha<1/2$, so this result does not contradict the $t^\alpha$ behaviour predicted for $\alpha > \frac 12$ in Eq.~\eqref{eq:Jd_ag0.5}. 

\begin{figure}[t]
    \centering
   \includegraphics[width=12 cm]{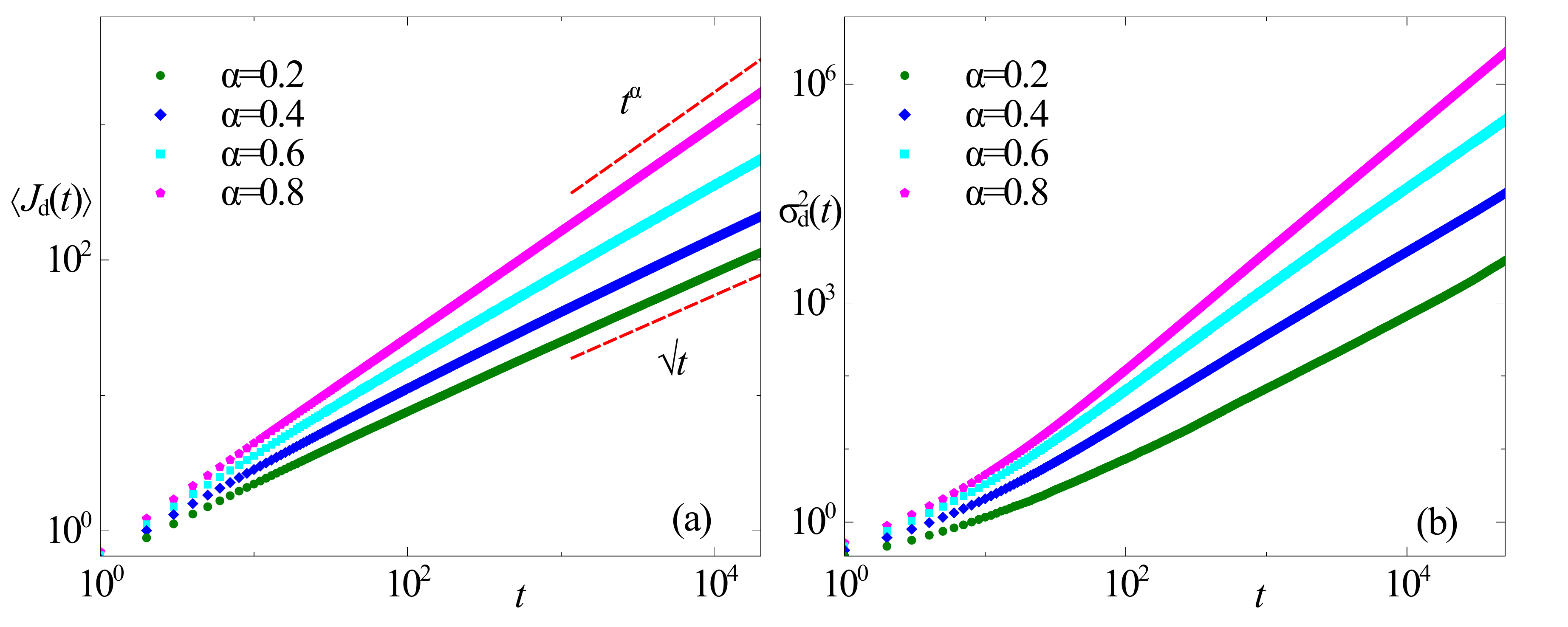}
    \caption{Behaviour of the mean and variance of the diffusive current $J_\text{d}(t)$ for $\alpha <1$: (a) Plot of $\la J_{\text d}(t)\ra$ {\it vs} $t$ for different values of $\alpha$, measured from numerical simulations. The upper red dashed line indicates the $t^\alpha$ growth, expected for $\alpha > 1/2$ [see Eq.~\eqref{eq:aghalf}], for the curve corresponding to $\alpha =0.8$. The lower red dashed curve indicates the $\sqrt{t}$ growth expected for $\alpha<1/2$ [see Eq.~\eqref{eq:alhalf}]. (b) Plot of the variance $\sigma_\text{d}^2(t)$ corresponding to the same values of $\alpha$ as in (a), showing the late-time algebraic growth. The corresponding exponents measured from these data are shown in Fig.~\ref{fig:exponents_moments}. We have used $L=1000$ and $t_0=0.01$ for the numerical simulations. }\label{fig:diffusive_current_al1_thermo}
\end{figure}

Combining Eqs. \eqref{eq:aghalf} and \eqref{eq:alhalf} we get the complete behaviour of the average diffusive current for $\alpha <1$ which is quoted in Eq.~\eqref{eq:Jdav_t}. Figure~\ref{fig:diffusive_current_al1_thermo}(a) shows the behaviour of $\la J_\text{d}(t) \ra$ for different values of $\alpha<1$, measured from numerical simulations which validates our analytical prediction.

\subsubsection{Average diffusive current for $\alpha>1$:}

For $\alpha >1$, the average diffusive current can again be obtained using the definition Eq.~\eqref{eq:Jd_jd} where the average instantaneous current $\la j_\text{d}(t) \ra$ is computed from the density profile. In the late-time regime, the density profile reaches a non-trivial stationary profile, and the average instantaneous current reaches a stationary value as well [see Eq \eqref{eq: avg_j}],
\bea
\la j_\text{d} \ra = \frac{2}{L}\sum_{n=1,3,5...}^{L-1} G(\lambda_n).
\eea
Consequently, in the long-time regime, we expect that the average diffusive current will grow linearly with time $t$. In fact, from Eqs. \eqref{eq: stat_density_profile_ag1}and \eqref{eq: avg_j}, we expect, in the thermodynamic limit,
\bea    \label{eq:jdag1_linear}
\la J_\text{d}(t) \ra \simeq  \frac{(\alpha-1) t}{\alpha t_0} \int_0^{2\pi} \frac{dq}{2\pi} \left( \frac{1-e^{-\lambda_q t_0}}{\lambda_q} + t_0 E_\alpha(\lambda_q t_0) \right). 
\eea
where, as before, $\lambda_q=2(1-\cos q)$ and $E_\alpha(z)$ denotes the exponential integral function. The $q$-integral is hard to evaluate analytically for arbitrary $t_0$. However, it is easy to see that the integrand is well behaved and hence, it converges for all $t_0$, implying   $\la J_\text{d}(t) \ra \propto t$ for $\alpha >1$.  Figure \ref{fig:Jav_J2av_ag1}(a) shows a plot of $\la J_\text{d}(t) \ra$ for different values of $\alpha >1$ which illustrates this linear growth. In fact,  for small $t_0$ we can find the leading order contribution  by expanding the $E_\alpha(z)$ for small $z$ which  leads to, 
\bea    
\la J_\text{d}(t) \ra \simeq  t  \left[ 1 - (4t_0)^{\alpha -1}\, \frac{\Gamma(2-\alpha)\Gamma(\alpha - \frac 12)}{\sqrt \pi \, \Gamma(1+\alpha)}\right]. \label{eq:agone}
\eea

To summarize, in the presence of the non-Markovian resetting, the average diffusive current for SEP shows non-trivial algebraic growth $\la J\text{d}(t) \ra \sim t^{\theta(\alpha)}$ at late times $t \gg t_0$. The dependence of the growth exponent $\theta(\alpha)$ on $\alpha$ is quoted in Eq.~\eqref{eq:Jdav_t}.

\subsection{Variance of the diffusive current}

\begin{figure}[t]
    \centering
    \includegraphics[width=13cm]{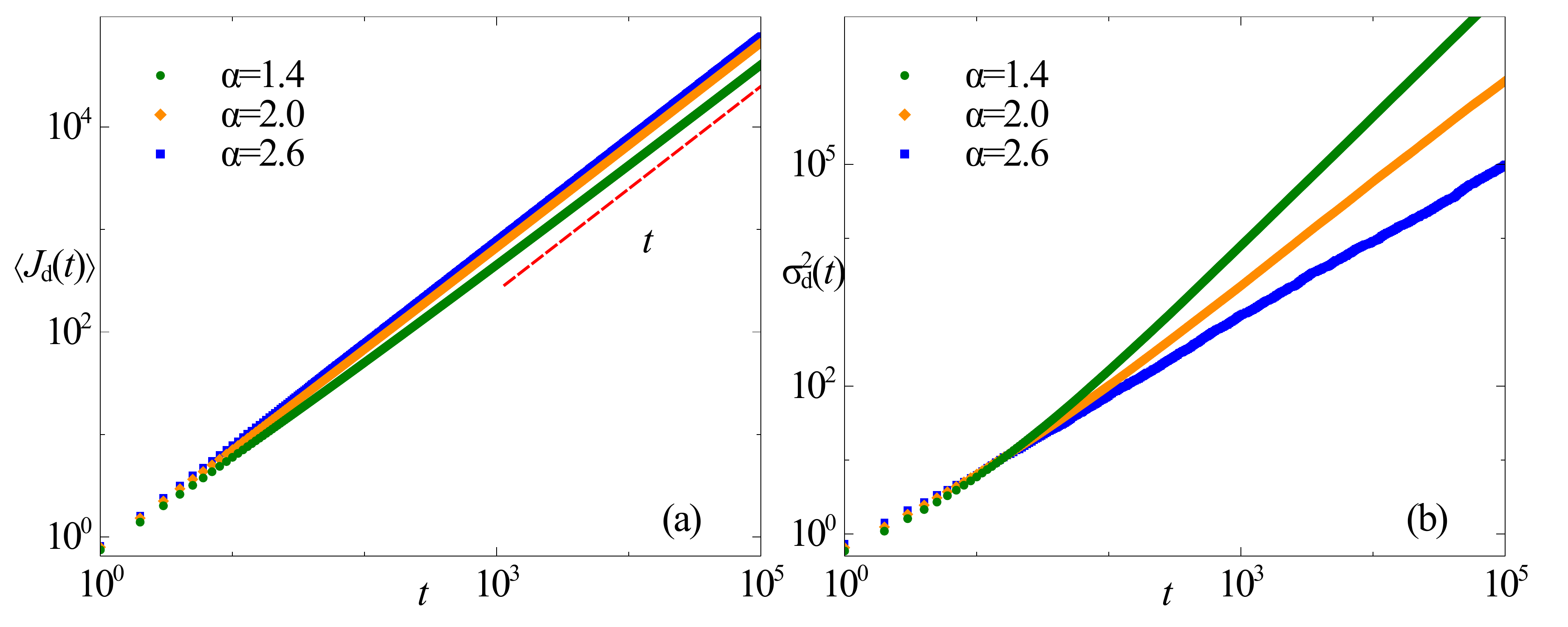}
    \caption{Behaviour of the mean and variance of the diffusive current $J_\text{d}(t)$ for $\alpha >1$: (a) Plot of $ \la J_\text{d}(t)\ra$ {\it vs} $t$ for different values of $\alpha >1$; the red dashed line indicates the expected linear growth [see Eq.~\eqref{eq:jdag1_linear}]. (b) Plot of the variance $\sigma_\text{d}^2(t)$ of the same of data. The simulations are performed on a lattice of size $L=1000$ with $t_0=0.1$.}\label{fig:Jav_J2av_ag1}
\end{figure}

It is also interesting to investigate the behaviour of the variance of the diffusive current
\bea
\sigma_\text{d}^2(t) \equiv \la J_\text{d}^2(t) \ra - \la J_\text{d}(t) \ra^2 
\eea
in the presence of the non-Markovian resetting protocol. We use numerical simulations to measure $\sigma_\text{d}^2(t)$ for different values of $\alpha$. Figures \ref{fig:diffusive_current_al1_thermo}(b) and \ref{fig:Jav_J2av_ag1}(b) show plots of $\sigma_\text{d}^2(t)$ as a function of time $t$ for $\alpha <1$ and $\alpha >1$, respectively, which indicate an algebraic growth $\sigma_\text{d}^2(t) \sim t^{\zeta(\alpha)}$ for the variance as well. Although we do not have any analytical predictions for this, the numerically estimated exponent $\zeta(\alpha)$ is plotted in Fig \ref{fig:exponents_moments} as a function of $\alpha$ which seems to show a non-trivial behaviour.

\subsection{Probability distribution of the diffusive current}

To further characterise the fluctuations of the diffusive current, we next look at the probability distribution $P(J\text{d},t)$, which denotes the probability that the time-integrated diffusive current is $J\text{d}$ after a time interval $t$. The renewal method used to compute the average diffusive current is not directly applicable here, and instead, we use the trajectory-based approach introduced in Sec. \ref{sec:RnTb}. 

Once again, we consider a trajectory of the system over a time-interval $[0,t]$, with $n$ resetting events.  The time-integrated diffusive current for such a trajectory can be expressed as 
\bea
J_\text{d} = \sum_{i=1}^n J_i(\tau_i),
\eea
where $J_i(\tau_i)$ denotes the time-integrated diffusive current during $\tau_i$, which denotes the interval between the $i$-th and $(i-1)$-th resetting events. As before, we must have $t = \sum_{i=1}^n \tau_i$. Let us recall that, between 
two consecutive resetting events the configuration of the system independently evolves following ordinary SEP dynamics. Then the probability density $p_n(J\text{d},t)$ of the diffusive current being $J\text{d}$, given exactly $n$ resetting events take place within time $t$, can then be expressed as,
\bea
\fl p_n(J_\text{d},t) =\left \{
\begin{split}
& f(t) P_0(J_\text{d},t),  \qquad \qquad \qquad \qquad  \qquad \qquad \qquad  \text{for} ~ n=0 \cr
 &\int \prod_{i=1}^{n} d \tau_i\,d \tau_{n+1} \psi(\tau_i) f(\tau_{n+1}) P_0(J_i,\tau_i) P_0(J_{n+1},\tau_{n+1}) \cr
 & \qquad \qquad \times \delta \left(J_\text{d} - \sum_{i=1}^{n+1} J_i\right)  \delta \left(t-\sum_{i=1}^{n+1} \tau_i \right), \quad  \text{for} ~ n>0
 \end{split}
 \right. \label{eq:Jpn}
\eea
where $P_0(J,\tau)$ denotes the probability of the diffusive current being $J$ in an interval $\tau$, in the absence of resetting.  The complete probability distribution of $J_\text{d}$, in the presence of resetting, then can be formally expressed by combining the contribution of all possible trajectories, 
\bea
P(J_\text{d},t) = \sum_{n=0}^\infty p_n(J_\text{d},t). \label{eq:Jsum}
\eea

Of course, it is rather hard to evaluate $P(J_\text{d},t)$ exactly for all times from the above infinite sum. However, at short times when the typical number of resetting events is expected to be small, considering the first few terms in the expansion should suffice. Thus, in this regime, one can use Eq.~\eqref{eq:Jsum} as a perturbative series in $n$ to compute $P(J\text{d},t)$ at short times. 

\begin{figure}[t]
    \centering
    \includegraphics[width=15cm]{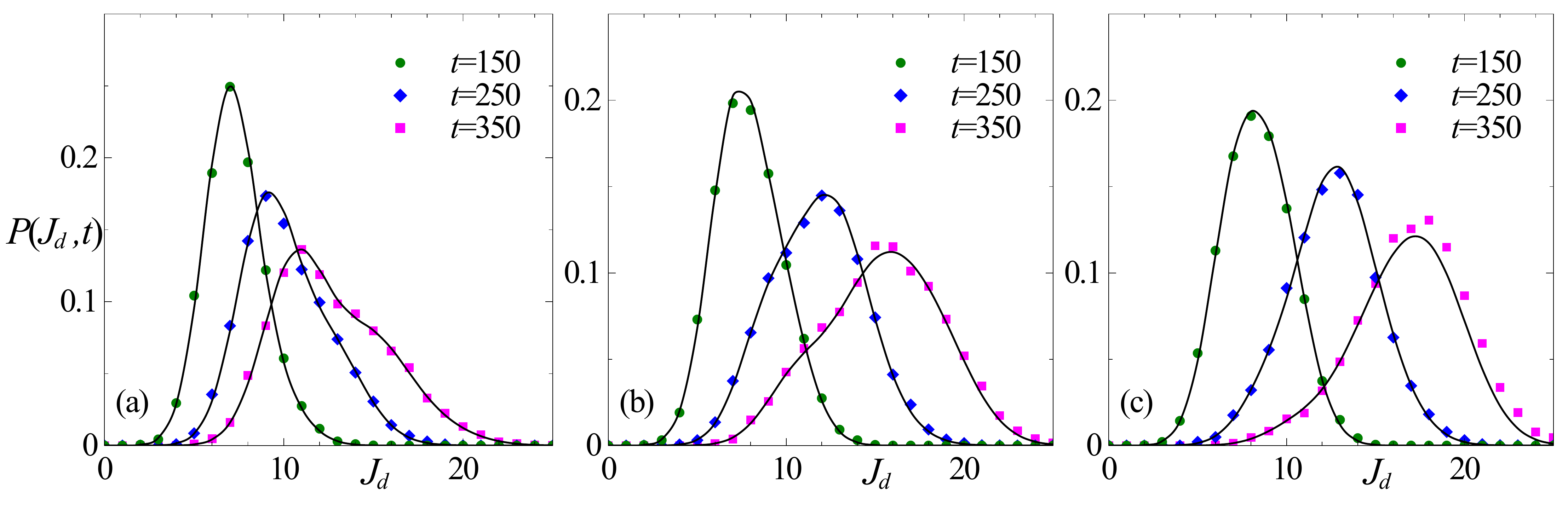}
    \caption{Short-time distribution of diffusive current for $t_0=100$ and different small values of $t (\gtrsim t_0)$ for $\alpha=0.5$ (a), $\alpha=1.5$ (b) and $\alpha=2.5$ (c). The symbols show the data obtained from numerical simulations and the solid lines show the analytical predictions from the perturbative approach Eq. \eqref{eq:Jsum}, computed up to $n=2$. The simulations are done on a lattice of size $L=1000$.}
    \label{fig: Jd_dist_pert_compare}
\end{figure}

In the following we show the explicit computations for $p_n(J_\text{d},t)$ up to the $n=2$ term, which provides a good approximation for the distribution in the $ t \gtrsim t_0$ regime [see Fig \ref{fig: Jd_dist_pert_compare}]. Of course, for $n=0$, i.e., when no resetting occurs, the system evolves following ordinary SEP dynamics, and the diffusive current distribution is given by [see Eq. \eqref{eq:P0}],
\bea
p_0(J_\text{d},t) = f(t) P_0(J_\text{d}, t)\label{eq:p0}
\eea
where $f(t)$ [see Eq. \eqref{eq:f}] denotes the probability that no resetting has occurred until time $t$. The contribution for $n=1$, i.e., trajectories with one resetting event can also be written following Eq. \eqref{eq:Jpn}, and is given by,
\bea
p_1(J_\text{d},t) &=  \int_0^t dt_1~ \psi(t_1)f(t-t_1)\int dj_1~ P_0(j_1, t_1) P_0(J_\text{d}-j_1, t-t_1)
\eea
which can be evaluated numerically using the Gaussian form of $P_0(J_0,t)$ given in Eq.~\eqref{eq:P0}. Similarly, the contribution from $n=2$ trajectories, 
\bea
p_2(J_\text{d},t) &= \int_0^t ~dt_1 \int_0^{t-t_1} dt_2~  \psi(t_1)\psi(t_2) f(t-t_1-t_2) \times \\ 
& \quad \int dj_1 \int dj_2 ~ P_0(j_1,t_1)P_0(j_2,t_2)P_0(J_\text{d}-j_1-j_2,t-t_1-t_2)
\eea
can also be evaluated numerically.

Figure~\ref{fig: Jd_dist_pert_compare} compares the diffusive current distribution, obtained from numerical simulations with different values of $\alpha$, with the analytical prediction given in Eqs.~\eqref{eq:Jpn}-\eqref{eq:Jsum}, evaluated up to $n=2$. For each value of $\alpha$, we measure the distribution $P(J_\text{d},t)$ for three different values of $t$:  (i) $t < 2t_0$, where there is at most one resetting event, and hence \eqref{eq:Jsum}, evaluated up to $n=1$ terms gives the exact distribution, (ii) $ 2 t_0 < t < 3 t_0 $ where there can be at most two resetting events, and summing up to $n=2$ gives $P(J_\text{d},t)$  exactly, and (iii) $t > 3 t_0 $ where finite contributions from $n >2$ are expected. As shown in Fig.~\ref{fig: Jd_dist_pert_compare}, the analytical predictions work quite well even for $t > 3t_0$, validating our perturbative procedure. For all values of $\alpha$, the distribution has a single peak which shifts towards larger $J_\text{d}$ value as $t$ increases. However, the shape of the distribution changes with $\alpha$ --- while positively skewed for smaller $\alpha$, the distribution turns negatively skewed as $\alpha$ is increased.

\begin{figure}[t]
        \centering
        \includegraphics[width=15cm]{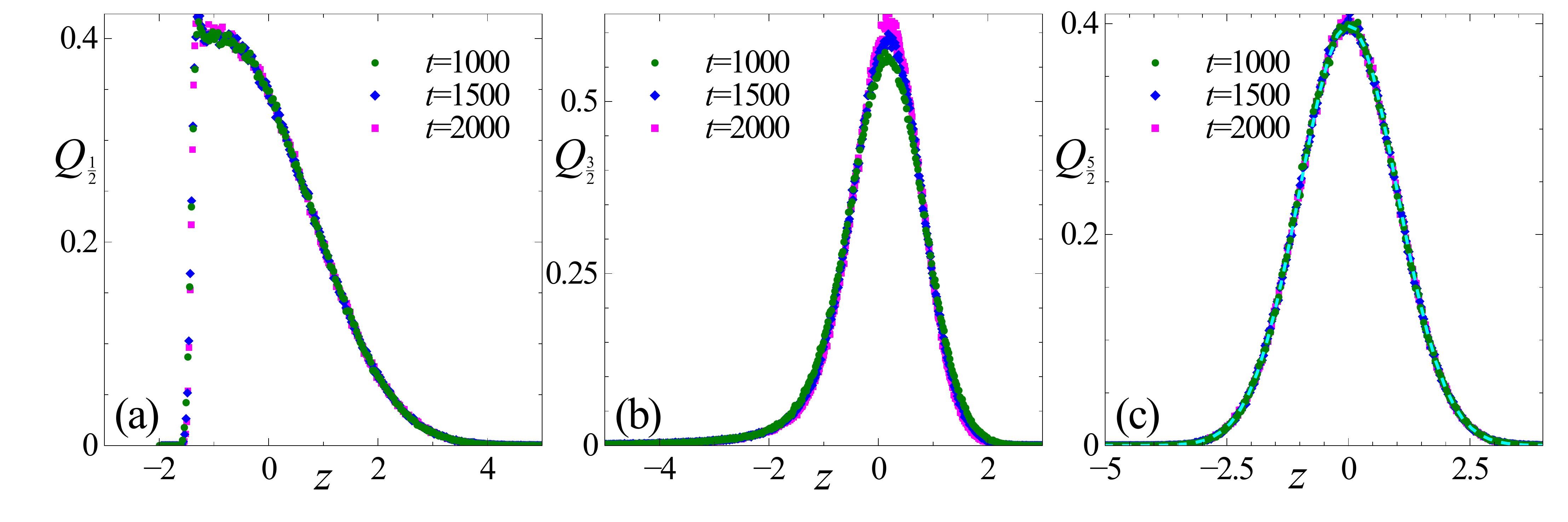}
        \caption[Diffusive current distributions for $\alpha>1$.]{The scaling collapses of the diffusive current distribution $P(J_\text{d},t)$, following Eq.~\eqref{eq:scaling_PJd} for  $\alpha=\frac 12$ (a), $\alpha=\frac 32$ (b), and $\alpha=\frac 52$ (c).  In each case, the distribution of the scaled variable $z = (J_\text{d} - \la J_\text{d}(t) \ra)/\sigma_\text{d}(t)$,  obtained from numerical simulations,  is plotted for different (large) values of $t$, which collapse onto a single curve. The three panels show the qualitatively different shapes of the scaling function ${\cal Q}_\alpha(z)$. For $\alpha >2$ (c), the scaling function approaches the standard normal distribution, which is shown as a dashed line. 
        The simulations are done on a lattice of size $L=1000$ and using $t_0=0.01$.}
        \label{fig:Jd_density_scale}
    \end{figure}

At long-times $t \gg t_0$, however, the perturbative procedure is not expected to work since the typical number of resetting events is large in this regime. We take recourse to numerical simulations in this regime and find that the diffusive current distribution in the long-time regime has $\alpha-$dependent shapes. In fact, based on the numerical simulations, we propose the following scaling form for the diffusive current distribution in the long-time regime, 
\bea    
P(J_\text{d},t) = \frac 1{\sigma_\text{d}(t)} {\cal Q}_{\alpha}\left( \frac{J_\text{d} - \la J_\text{d}(t) \ra  }{\sigma_\text{d}(t)} \right) \label{eq:scaling_PJd}
\eea
where $ \la J_\text{d}(t)\ra$ and $\sigma_\text{d}(t)$ denote the mean and variance of $J_\text{d}$, respectively [see Figs.~\ref{fig:diffusive_current_al1_thermo}-\ref{fig:Jav_J2av_ag1}].  Figure~\ref{fig:Jd_density_scale} shows the scaling collapse of the measured distribution according to \eqref{eq:scaling_PJd} for three values of $\alpha$, which validates our conjecture. Interestingly once again, the shape of the scaling function is positively skewed for smaller $\alpha$, turns negatively skewed as $\alpha$ is increased, before becoming symmetric for even larger $\alpha$.  In fact, for $\alpha >2$, the scaling function approaches a Gaussian form [see Fig.~\ref{fig:Jd_density_scale}(c)] which can be heuristically understood from a central-limit like arguments.

\section{Total current}\label{sec:Total current}

The total current $J(t)$, as defined previously, measures the total number of particles crossing the central bond towards the right. Clearly, each resetting event also resets the total current to zero, since all the particles are put back to the left half of the lattice at each resetting.  Consequently, the distribution of the total current $\mathbb{P}(J,t)$ must satisfy the renewal equation similar to the configuration probability [see Eq.~\eqref{eq: nonm_renewal_F}]
\bea
\mathbb{P}(J, t) = \int_0^t d \tau \, F(\tau, t-\tau)P_0(J,\tau),
\label{eq:renewal_PJt}
\eea
where $P_0(J,t)$ denotes the diffusive current distribution in the absence of the resetting.  It is also useful to write the renewal equation in the Laplace space [see Eq. \eqref{eq:Pcs}],
\bea
\tilde{\mathbb{P}}(J, s) \equiv \int_0^\infty dt\, e^{-s t } \, {\mathbb{P}}(J, t) = \frac 1{1 - \tilde \psi(s)} \int_0^\infty d \tau ~e^{-s \tau} f(\tau)  P_0(J,\tau)\label{eq:renewal_PJs}
\eea
where, as before, $\tilde \psi(s)$ denotes the Laplace transform of the waiting time distribution $\psi(\tau)$. Similar to the configuration probability,  $\mathbb{P}(J, t)$ also reaches a stationary limit for $\alpha > 1$, which is formally given by,
\bea
\mathbb{P}_\text{st}(J) = \frac 1{\la \tau\ra} \int_0^\infty d\tau \, f(\tau) P_0(J,\tau). \label{eq:Pst_Jtot}
\eea
In the following, we investigate the behaviour of the total current fluctuations.

\subsection{Average total current}

We start by computing the average total current $\la J(t) \ra$, which is most conveniently using the renewal equation in the Laplace space. Multiplying \eqref{eq:renewal_PJs} by $J$ and integrating over $J$, we get,
\bea    
\mathbb{\tilde J}(s) &=&  \frac 1{1- \tilde \psi(s)} \int_0^\infty d\tau~ e^{-s \tau} f(\tau) \la J_0(\tau) \ra, \label{eq:trajectorybased_total_current}
\eea
where, $\mathbb{\tilde J}(s) = \int_0^\infty dt~e^{-st} \la J(t) \ra$ denotes the Laplace transform of the average total current, and $\la J_0(\tau) \ra$ is the average current in the absence of resetting. We are primarily interested in the long-time behaviour of $\la J(t) \ra$, which is controlled by the small $s$ behaviour of the $\mathbb{\tilde J}(s)$. For small $s$, the integral in \eqref{eq:trajectorybased_total_current} is dominated by the large $\tau$ behaviour of the integrand and it suffices to consider the large-time behaviour of $\la J_0(\tau) \ra \simeq \sqrt{\tau/\pi}$ [see Eq.~\eqref{eq:mean_n_var}], which yields,

\bea    
\mathbb{\tilde J}(s) \simeq \frac{1}{1-\tilde \psi(s)}\left(
    \frac{1}{2 s^{3/2}} - \frac{t_0^{3/2}}{\sqrt{\pi}}E_{-\frac 12}( st_0) + \frac{t_0^{3/2}}{\sqrt{\pi}}E_{\alpha-\frac 12}( st_0) \right), \label{eq:js_expression}
\eea

where $E_\nu(z)$ denotes the exponential integral function (see Sec. 8.19 in  \cite{DLMF}). To obtain the large-time behaviour of $ \la J(t) \ra$ it suffices to consider the leading order behaviour of $\mathbb{J}(s)$ for $ s \to 0$ and then invert the Laplace transform. The details of this computation is provided in the \ref{app: trajectorybased_avg_jt}, here we quote the main results. It turns out that the large-time behaviour of the average total current also depends strongly on the value of $\alpha$. For $\alpha \le 1$, we have, for small $s$,
\bea
\mathbb{\tilde J}(s) = \frac{\Gamma(3/2-\alpha)}{\Gamma(1-\alpha)\sqrt{\pi}}s^{-3/2} + O(s^{-\alpha}), 
\eea
which, in turn, leads to,
\bea    
\la J(t) \ra \simeq  \frac{2\Gamma(3/2-\alpha)}{\Gamma(1-\alpha)\pi}\, \sqrt{t},  \label{eq:Jav_al1}
\eea
in the late-time regime. On the other hand, for $1< \alpha \le 3/2$, we get,
\bea    
\mathbb{\tilde J}(s) =  \frac{\alpha-1}{\alpha} t_0^{\alpha-1} \Gamma(3/2-\alpha) s^{\alpha-5/2} + O(s^{-3/2}), 
\eea
which leads to an algebraic late-time growth with an $\alpha$--dependent exponent,
\bea    \label{eq:Jav_al3b2}
\la J(t) \ra \simeq \frac{2 (\alpha-1) \, t_0^{\alpha -1}}{\sqrt{\pi}\alpha(3-2 \alpha)}   \, t^{3/2-\alpha}.
 \label{eq:Jav_ag1}
\eea

\begin{figure}[t]
    \centering
    \includegraphics[width=17 cm]{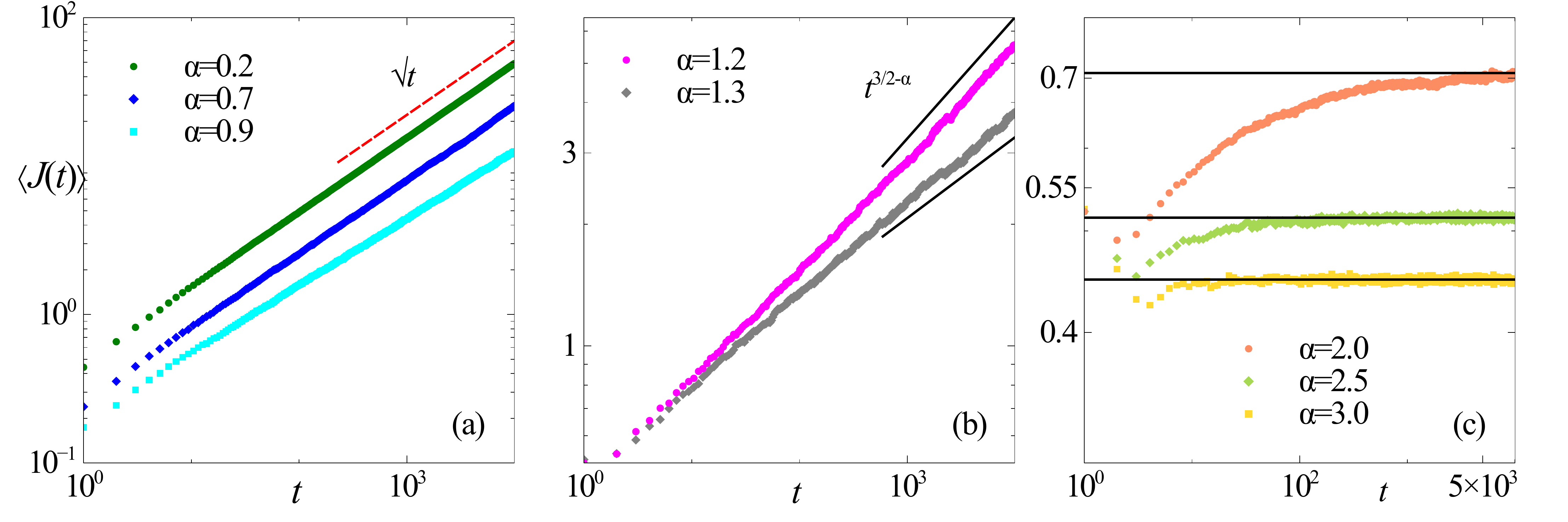}
    \caption{Plot of the average total current $\la J(t)\ra$ versus $t$ for $\alpha <1$ (a), $1 < \alpha < 3/2$, (b) and $\alpha > 3/2$ (c). The red dashed line in (a) indicates the $\sqrt{t}$ growth for $\alpha<1$ [see Eq. \eqref{eq:Jav_al1}], and the solid black lines in (b) indicate the predicted $t^{3/2-\alpha}$ growth for $1 < \alpha < 3/2$  [see Eq. \eqref{eq:Jav_al3b2}]. The black solid lines in (c) indicate the predicted stationary values [see Eq. \eqref{eq:Jtstat}]. The system size $L=1000$ for all three panels while $t_0=0.01$ for (a) and $t_0=1$ for (b) and (c). }
    \label{fig:Jtmom_avg}
\end{figure}

Combining Eqs.~\eqref{eq:Jav_al1} and \eqref{eq:Jav_ag1}, we get Eq.~\eqref{eq:Jav_growth} which has been quoted earlier. Figure~\ref{fig:Jtmom_avg} (a) and (b) show plots of $\la J(t) \ra$ for different values of $\alpha < 3/2$ which confirms this prediction.

For $\alpha > 3/2$, $\mathbb{\tilde J }(s) \sim 1/s$ for small $s$, indicating that the average total current eventually reaches a stationary value  $J_\text{st}$  in this regime. This stationary value can be approximately determined from the coefficient of $1/s$ in the series expansion of $\mathbb{\tilde J }(s) $ in Eq.~\eqref{eq:js_expression}. However, this approximate estimate of $J_\text{st}$ is not very accurate since for large $\alpha > 3/2$ the typical time between consecutive resets becomes small enough so that the behaviour of the current at shorter times becomes relevant.  Instead, we compute $J_\text{st}$ exactly using Eq.~\eqref{eq:Pst_Jtot} --- multiplying both sides of the equation by $J$ and integrating over $J$, we get,
\bea 
J_\text{st} =  \frac 1{\la \tau \ra} \int_0^\infty d\tau f(\tau)\la J_0(\tau) \ra. \label{eq:Jt_stationary}
\eea
Using the exact expression for $\la J_0(t) \ra$ from Eq.~\eqref{eq:mean_n_var}, we get,
\bea
\fl \qquad J_\text{st}  = \frac{\alpha-1}{\alpha}\Bigg[ \frac{1}{6} e^{-2t_0} \Big(4t_0I_0(2t_0) + (4t_0-1)I_1(2t_0)\Big ) + (4t_0)^{\alpha-1}\frac{\Gamma(1-\alpha)\Gamma(\alpha - \frac 32)}{4\sqrt{\pi}\Gamma( \alpha)} \cr
~~~ - \frac{t_0}{2-\alpha}  {}_2F_2 \left(\frac 12, 2-\alpha ; 2, 3-\alpha; - 4 t_0 \right) \Bigg ], \label{eq:Jtstat}
\eea
where $ _2F_2(a_1,a_2;b_1,b_2;z)$ is a generalised hypergeometric function (see Sec.~16.2 of \cite{DLMF}).
Figure~\ref{fig:Jtmom_avg}(c) shows a plot of $\la J(t) \ra$ versus $t$ for $\alpha > 3/2$; the eventual stationary values agree perfectly with the above prediction. As expected, the time required to reach the stationary value decreases as $\alpha$ is increased.

\subsection{Variance of the total current}

It is also interesting to look at the variance of the total current, 
\bea
\sigma^2(t) = \la J^2(t) \ra - \la J(t) \ra^2.
\eea
The second moment $\la J^2(t) \ra$ can be computed following the same trajectory-based approach used in the previous section. From \eqref{eq:renewal_PJs}, we get its Laplace transform,
\bea
\mathbb{\tilde J}_2(s) &=& \int_0^\infty dt~e^{-st} \la J^2(t) \ra 
= \frac 1{1- \tilde \psi(s)} \int_0^\infty d\tau~ e^{-s \tau} f(\tau) \la J_0^2(\tau) \ra. \label{eq:trajectorybased_ttoal_current_var}
\eea
Once again, we focus on the small $s$ behaviour, which can be computed by using the long-time expression of $\la J^2_0(t) \ra$ from \eqref{eq:mean_n_var}, and is given by,
\bea
\mathbb{\tilde J}_2(s) = \frac 1{1- \tilde \psi(s)} \Big\{ \frac {1}{\pi s^2} \Big[ 1-e^{-st_0}(1+st_0) + \frac{\pi}{2}\left(1- \frac 1{\sqrt{2}}\right)\sqrt{s} \Big ] \cr
\qquad + \frac{t_0^2}{\pi} E_{\alpha-1}(st_0)  
+ \frac{ t_0^{3/2} }{\sqrt{\pi}}  \left(1- \frac 1{\sqrt{2}}\right) \Big[ E_{\alpha- \frac 12}(st_0) -  E_{-\frac{1}{2}}(st_0) \Big] \Big\}. \label{eq:J2s}
\eea

\begin{figure}[t]
    \centering
    \includegraphics[width=12 cm]{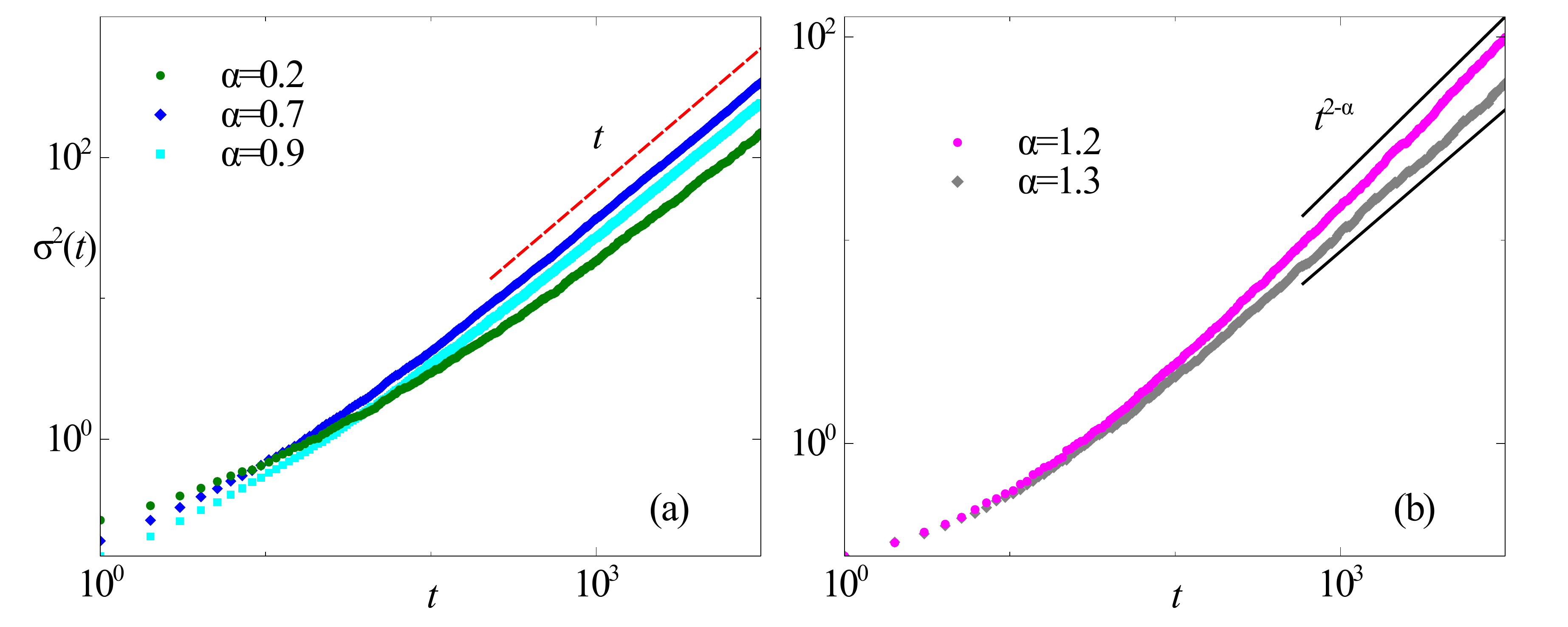}
    \caption{Plot of the variance of the total current $\sigma^2(t)$ versus $t$ for $\alpha <1$ (a) and $1 <\alpha < 2$ (b). The red dashed line in (a) indicates the  linear growth while the black solid lines  in (b) indicate the predicted $t^{2-\alpha}$ growth [see Eq.~\eqref{eq:C1}]. The numerical simulations have been done on a lattice of size $L=1000$, with $t_0=0.01$ for (a) and $t_0=1$ for (b).}
    \label{fig:Jt_var}
\end{figure}

The large time behaviour of $\la J^2(t) \ra$ can then be obtained from the $s \to 0$ behaviour of $\mathbb{\tilde J}_2(s)$ and then taking the inverse Laplace transform; the details of this computation is given in  \ref{app: trajectorybased_var_jt}.  Finally, for $\alpha <2$, we get an algebraic growth for the variance,  $\sigma^2(t) \sim t^{\Delta(\alpha)}$ with, 
\bea
 \Delta(\alpha)  = \left \{ \begin{split}
  1 \quad & ~~ \text{for} ~ \alpha <  1 \cr
  2-\alpha & ~~ \text{for} ~ 1 < \alpha \le 2. 
  \end{split}
  \label{eq:C1}
  \right.
\eea
Figure~\ref{fig:Jt_var} shows a plot of $\sigma^2(t)$ vs $t$ for different values of $\alpha$ which verifies this predicted algebraic growth. 

For $\alpha>2$, the variance approaches a stationary value at large times. However, it is difficult to calculate this stationary values as the complete time-dependent behaviour of $\la J_0(t)^2 \ra$ is not known.

\subsection{Distribution of the total current}

The long-time behaviour of the first two moments does not give any information about the shape of the total current distribution, nor on the large fluctuations. In this section, we use the renewal equation \eqref{eq:renewal_PJt} to characterize the late-time probability distribution of the total current $\mathbb{P}(J,t)$, for $\alpha <1$ and $\alpha >1$.

To obtain the distribution of the total current for  $\alpha <1$, we use the approximate form of $F(\tau, t-\tau)$ \eqref{eq: F_al1} in Eq.~\eqref{eq:renewal_PJt} along with the Gaussian form of $P_0(J,t)$ given in Eq.~\eqref{eq:P0}. The resulting integral cannot be performed analytically and we take resort to numerical integration.

\begin{figure}[th]
    \centering
    \includegraphics[width=12.5 cm]{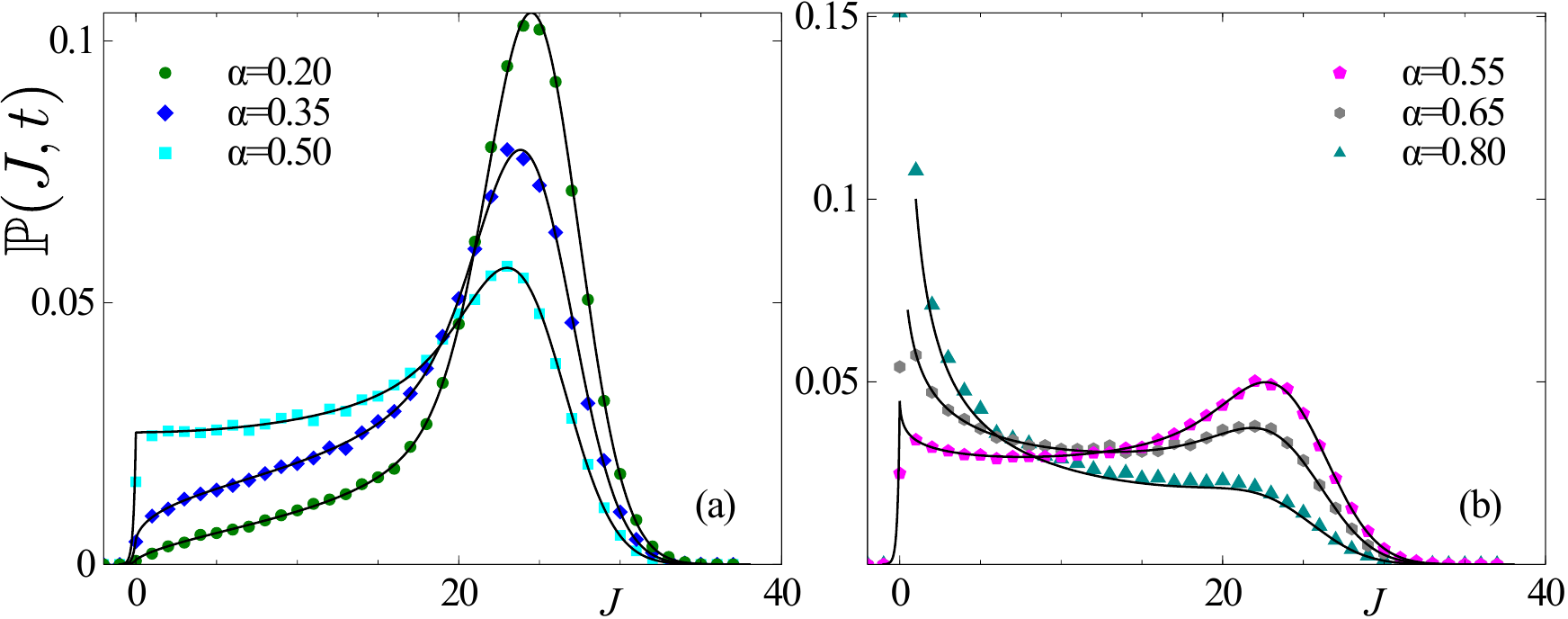}
    \caption{The total current distribution $\mathbb{P}(J,t)$ for a set of values of $\alpha<1$ at $t=2000$. For better visual clarity, we have shown the plots for $\alpha \le 1/2$ in (a) and $\alpha > 1/2$ in (b). The symbols indicate data from numerical simulations, which aredone with $L=1000$ and $t_0=0.01$, whereas the solid lines indicate the theoretical predictions from Eq. \eqref{eq:renewal_PJt}. }
    \label{fig:Jt_dist_al1}
\end{figure}

Figure~\ref{fig:Jt_dist_al1} shows the plot of the numerically evaluated $\mathbb{P}(J,t)$ for different values of $\alpha$, along with the same obtained from simulations. The total current distribution is strongly non-Gaussian with an asymmetric peak at a non-zero value of $J$. A qualitative difference appears between the shape of the distribution for $\alpha \le 3/4$ and $\alpha > 3/4$ --- in the former case the value of $\mathbb{P}(J,t)$ remains finite at $J=0$ while it diverges in the latter case.

\begin{figure}[t]
    \centering
    \includegraphics[width=10cm]{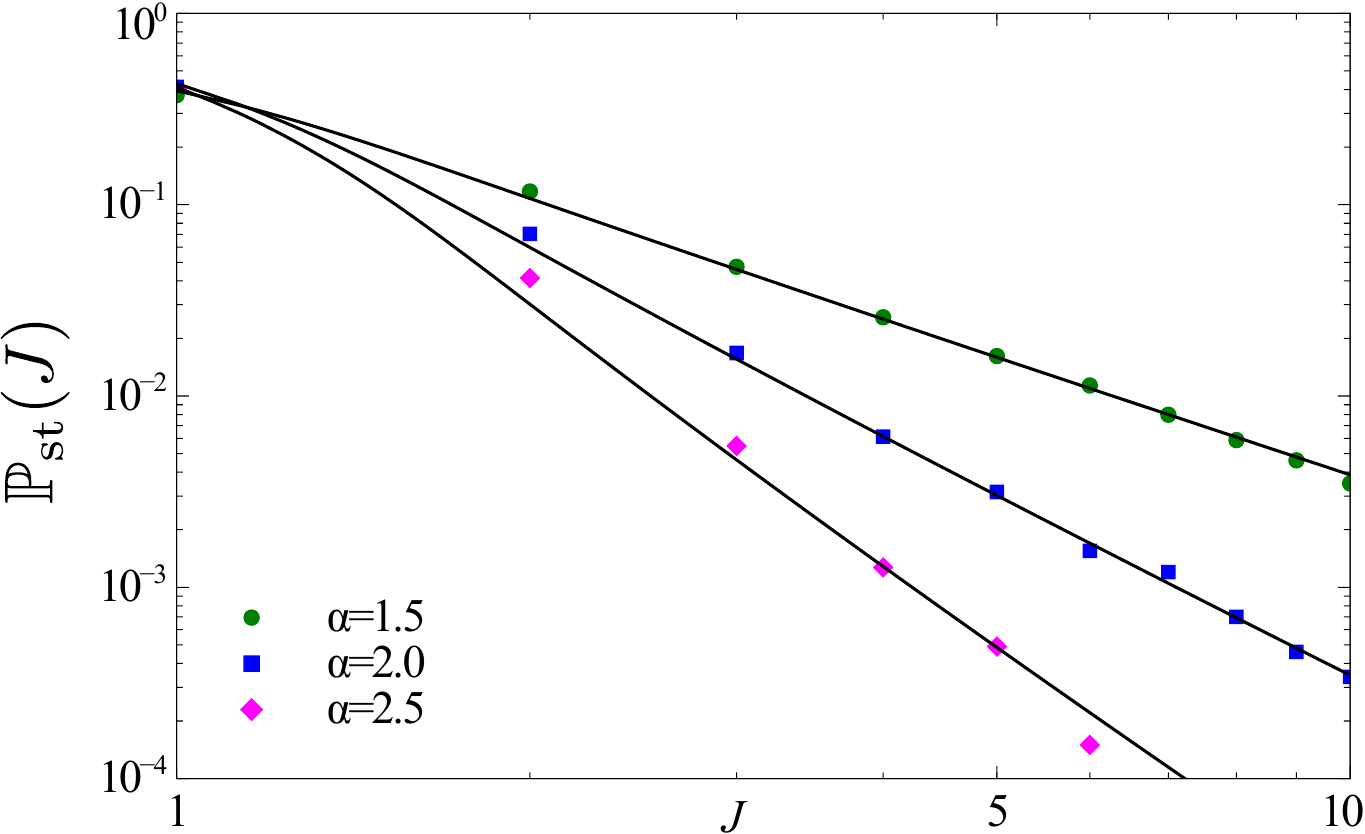}
    \caption{Plot of the stationary total current distribution for $\alpha>1$. The symbols indicate the data obtained from numerical simulations and the black solid lines indicate the analytical prediction from Eq.~\eqref{eq: Jt_distribution_ag1}. The simulations are performed on a lattice of size $L=1000$ with $t_0=1$.}
    \label{fig: Jt_distribution_ag1}
\end{figure}

For $\alpha>1$, the total current approaches a stationary distribution given by Eq.~\eqref{eq:Pst_Jtot},
\bea
\mathbb{P}_\text{st}(J) = \frac {\alpha -1}{\alpha t_0} \int_0^\infty d\tau \, f(\tau) P_0(J,\tau),\label{eq: Jt_distribution_ag1}
\eea
where $P_0(J,\tau)$ is the current distribution in the absence of resetting [see \eqref{eq:P0}]. Although the integral in the above equation cannot be performed analytically, the stationary distribution of the total current $\mathbb{P}_\text{st}(J)$ can be obtained for any value of $J$, using numerical integration. This is illustrated in Fig.~\ref{fig: Jt_distribution_ag1}, where where the predicted $\mathbb{P}_\text{st}(J)$ is compared with the same obtained from numerical simulations, showing a very good agreement.

\section{Conclusions}\label{sec:conclusion}

In this article, we explore the effects of a (non-Markovian) power-law stochastic resetting protocol on a simple model of interacting particles, namely the symmetric exclusion process. We consider a SEP on a half-filled periodic lattice, where, apart from the usual symmetric hopping dynamics, the system is reset to a step-like configuration, with all the particles clustered in the left-half of the system, after random waiting times, which are drawn from a power-law distribution. We show that, such a non-Markovian resetting protocol results in a range of intriguing features, absent in the Markovian resetting scenario. In particular, we focus on the density profile and particle currents, which show very different behaviour depending on the power-law exponent $\alpha$. 

We start by re-deriving the renewal equation for non-Markovian resetting in the configuration space using a trajectory based approach. We use this renewal equation to analytically compute the density profile. In particular, for $\alpha > 1$, the density reaches a non-trivial stationary profile, which we compute exactly. On the other hand, for $\alpha <1$, the density profile eventually becomes uniform. In this case, we analytically study the relaxation behaviour of the density profile.

The non-Markovian resetting protocol drastically changes the transport properties of the SEP. In the absence of the resetting, the diffusive particle current across the central bond grows $\sim \sqrt{t}$, while Markovian resetting with a constant rate gives rise to a linear temporal growth of the current. We show that,
in the presence of power-law resetting, the average diffusive current through a thermodynamically large system, shows an algebraic temporal growth at late-times $\sim t^{\theta(\alpha)}$ where, $\theta=1/2$ for $\alpha \le 1/2$, $\theta=\alpha$ for $1/2 < \alpha \le 1$ and $\theta=1$ for $\alpha>1$. We further show that the variance of the diffusive current also shows an algebraic growth with an $\alpha$-dependent exponent, which we characterize using numerical simulations. 

We also study the behaviour of the distribution of the diffusive current. Using the trajectory based approach perturbatively, we analytically obtain the distribution in the short-time regime, which shows strongly non-Gaussian and skewed features. Furthermore, we find that, the late-time distribution of the diffusive current follows a non-trivial scaling form [see Eq. \eqref{eq:scaling_PJd}], where the shape of the scaling function depends strongly on $\alpha$. It turns out that, for $\alpha >2 $ the scaling function approaches a Gaussian.

It is well known that introduction of dynamical resetting gives rise to a resetting current as well. We also study the fluctuations of the total current, which, again, shows much richer behaviour compared to the case of Markovian resetting. Using the renewal approach, we show that the average current also shows an algebraic temporal growth $\sim t^{\phi(\alpha)}$ with $\phi=1/2$ for $\alpha \le 1$, $\phi=3/2-\alpha$ for $1<\alpha \le 3/2$, and it approaches a stationary value for $\alpha>3/2$ which we compute exactly. The distribution of the total current can also be computed using the renewal approach. In particular, we find that, for $\alpha<1$ the distributions show strongly non-Gaussian fluctuations and some particularly interesting behaviour near $J=0$. For $\alpha>1$, we evaluate the stationary distributions using numerical integration, which also show strongly non-Gaussian typical fluctuations.

This work adds a significant contribution towards understanding the effect of non-Markovian resetting on transport properties of extended system using a simple model. It would be interesting to study how to behaviour of current changes for other non-Markovian and time-dependent resetting protocols.

\section{Acknowledgements}

U. B. acknowledges support from the Science and Engineering Research Board (SERB), India, under a Ramanujan Fellowship (Grant No. SB/S2/RJN-077/2018).

\appendix

\section{Computation of the late-time behaviour of the  average total current}    \label{app: trajectorybased_avg_jt}

In this Appendix we derive the large time growth exponents of the average total current $\la J(t) \ra$ using a trajectory based approach. First, we consider the large time approximation of the average diffusive current for SEP in absence of resetting \eqref{eq:mean_n_var}. Using $\la J_0(t) \ra = \sqrt{\frac{t}{\pi}}$ in Eq. \eqref{eq:trajectorybased_total_current}, we have
\bea    \label{eq:trajectorybased_total_current_appendix}
 \mathbb{\tilde J}(s) = \frac{1}{1-\tilde \psi(s)} \int_0^\infty ~d\tau ~e^{-s\tau} f(\tau) \sqrt{\frac{t}{\pi}}.
\eea
Here, $f(\tau)$ is the survival function as defined in \eqref{eq:f}, and $\tilde \psi(s)$ is the Laplace transform of $\psi(t)$. Using the explicit form of $f(\tau)$, the above integral can be performed exactly and leads to Eq.~\eqref{eq:js_expression} in the main text.

To extract the large-time behaviour of $\la J(t) \ra$, we investigate the small $s$ behaviour of $\mathbb{\tilde J}(s)$. To this end, we first note that, near $z=0$,
\bea
E_\nu(z) = \Gamma(1-\nu) z^{\nu -1}  + O(1).
\eea
Using the above equation and expanding the numerator and denominator of $\mathbb{\tilde J}(s)$ around $s=0$, we get,
\bea    \label{eq:series1}
\mathbb{\tilde J}(s) = \frac{1}{\sqrt{\pi}}\frac{c_0 s^{\alpha-3/2} + c_1 + c_2 s + {\cal O}(s^2)}{b_0 s^\alpha + b_1 s + {\cal O}(s^2)}.
\eea
with 
\bea
b_0 &=& t_0^\alpha \Gamma(1-\alpha), ~~ b_1=\frac{\alpha t_0}{\alpha-1}, \quad \text{and}, \cr
c_0 &=& t_0^\alpha \Gamma \left(\frac 32-\alpha \right), ~~ c_1 = \frac{4 \alpha t_0^{3/2}}{3(2\alpha-3)}, ~~ c_2 = \frac{4 \alpha t_0^{5/2}}{5(2\alpha-5)}
\eea
It is clear from the above equation that the leading order behaviour of $\mathbb{\tilde J}(s)$ for small $s$ depends on the value of $\alpha$. In the following, we discuss the different regimes separately.

\begin{itemize}

\item For $\alpha <1$, the small $s$-behaviour of  both the numerator and denominator in  Eq.~\eqref{eq:series1} are dominated by the respective first terms, which, in turn, gives,
\bea
\mathbb{\tilde J}(s) \simeq \frac{1}{\sqrt{\pi}}\frac{c_0}{b_0} s^{-3/2}.
\eea
Using the explicit forms of $c_0$ and $b_0$ and taking the inverse Laplace transform we get Eq. \eqref{eq:Jav_al1}, which describes the late-time growth of the average total current for $\alpha<1$. 

\item A different behaviour emerges when $1 < \alpha \le 3/2$. In this case, the numerator in Eq.~\eqref{eq:series1} is still dominated by the first term, while denominator is dominated by the $\sim s$ term. Hence, we have,
\bea
\mathbb{\tilde J}(s) = \frac{1}{\sqrt{\pi}}\frac{c_0}{b_1} s^{\alpha-5/2},
\eea
which, upon inverse Laplace transform leads to Eq.~\eqref{eq:Jav_al3b2} for the large-time behaviour in this regime. 

\item Finally, for $\alpha > 3/2$, we have, 
\bea    
\mathbb{\tilde J}(s) = \frac{1}{\sqrt{\pi}} \frac{c_1}{b_1 s},
\eea
which indicates that the average total current reaches a stationary value in this regime. However, simply taking the inverse Laplace transform of the above equation gives a rather inaccurate estimate of this stationary value, and we need to take recourse to a different method, as discussed in the main text.
\end{itemize}

\section{Computation of the variance of the  total current } \label{app: trajectorybased_var_jt}

Higher moments of the total current also follow the renewal equation \eqref{eq:renewal_PJt}, or equivalently, Eq.~\eqref{eq:renewal_PJs}. In particular, following Eq.~\eqref{eq:renewal_PJs} the Laplace transform of the second moment of the total current can be expressed as,
\bea 
\mathbb{\tilde J}_2(s) &=& \int_0^\infty dt~e^{-st} \la J^2(t) \ra = \frac 1{1- \tilde \psi(s)} \int_0^\infty d\tau~ e^{-s \tau} f(\tau) \la J_0^2(\tau) \ra. \label{eq:J2s_int}
\eea
The long-time behaviour of $\la J_0^2(\tau) \ra$, the second moment of the current in the absence of resetting, can be obtained from Eq.~\eqref{eq:mean_n_var},
\bea
\la J_0^2(t) \ra = \sigma^2_0(t) + \la J_0(t) \ra^2 \simeq \left(1-\frac{1}{\sqrt{2}}\right) \sqrt{\frac{t}{\pi}} +  \frac{t}{\pi}. \label{eq:Jsec}
\eea 
Using \eqref{eq:Jsec} in \eqref{eq:J2s_int}, and performing the integral we get Eq.~\eqref{eq:J2s} in the main text. 

To extract the large time behaviour of $\la J^2(t) \ra$ we proceed in the same way as before, by expanding the numerator and denominator around $s=0$,
\bea    \label{eq:series2}
\mathbb{\tilde J}_2(s) = \frac{d_0 s^{\alpha-2} + d_1 s^{\alpha-3/2} + d_2 + d_3 s + {\cal O}(s^2)}{b_0 s^\alpha + b_1 s + {\cal O}(s^2)},
\eea
where 
\bea
d_0 = \frac{t_0^\alpha \Gamma(2-\alpha)}{\pi},~~ d_1 =\left(1-\frac{1}{\sqrt{2}}\right) \sqrt{\pi} t_0^\alpha \Gamma(3/2-\alpha),\\ \n
d_2=\left(1-\frac{1}{\sqrt{2}} \right) \frac{4\sqrt{\pi} \alpha t_0^{3/2}}{3(2\alpha-3)} - \frac{\alpha t_0^2}{2\pi(2-\alpha)}, \quad \text{and,} \\  \n
d_3=-\left(1-\frac{1}{\sqrt{2}} \right) \frac{4\sqrt{\pi} \alpha t_0^{5/2}}{5(2\alpha-5)} + \frac{\alpha t_0^3}{3\pi(3-\alpha)}.
\eea
Clearly, the dominating behaviour of $\mathbb{\tilde J}_2(s)$ near $s=0$ depends on the value of $\alpha$ and we proceed in the same way as before, exploring the different regimes. 

\begin{itemize}

\item For $\alpha < 1$, both the numerator and denominator in \eqref{eq:series2} are dominated by the corresponding first terms, leading to, 
\bea
\mathbb{\tilde J}_2(s)\simeq \frac{d_0}{b_0} s^{-2} = \frac{(1-\alpha)}{\pi }s^{-2}.
\eea
Taking inverse Laplace transform, we get,
\bea
\la J^2(t) \ra \simeq \frac{1-\alpha}{\pi}t,
\eea
which leads to the variance 
\bea
\sigma^2(t) = \la J^2(t) \ra - \la J(t) \ra^2 \simeq \frac t \pi  \left(1-\alpha -\frac{4 \Gamma(3/2-\alpha)^2}{\pi\Gamma(1-\alpha)^2} \right),
\eea
where we have used Eq.~\eqref{eq:Jav_al1}. Clearly, the variance grows linearly in time for large times.

\item For $1<\alpha<2$, the leading order contribution in the numerator of Eq. \eqref{eq:series2} comes from the first term, while the denominator is dominated by the second term. This, in turn, leads to, 
\bea
\mathbb{\tilde J}_2(s) \simeq \frac{d_0}{b_1} s^{\alpha-3} = \frac{ t_0^{\alpha-1} (\alpha-1) \Gamma(2-\alpha)}{\alpha\pi} s^{\alpha-3}.
\eea
Consequently, we have, 
\bea
\la J^2(t) \ra = \frac{t_0^{\alpha-1}(\alpha-1)}{\pi \alpha (2-\alpha)}t^{2-\alpha}. 
\eea

In fact, in this regime, the variance $\sigma^2(t) = \la J^2(t) \ra - \la J(t) \ra^2$ also grows with the same exponent since $\la J(t) \ra^2 \sim t^{3-2\alpha}$ grows with a smaller exponent for $\alpha < 3/2$ and $\la J(t) \ra^2$ becomes stationary for $\alpha > 3/2$. 

\item For $\alpha>2$, we have,
 \bea
 \mathbb{\tilde J}_2(s) \simeq \frac{d_2}{b_1 s} ,
 \eea
 indicating that the second moment reaches a stationary value in this regime. However, similar to the case of average current, an inverse Laplace transform of the above equation does not provide a very accurate estimate of the stationary value. 
\end{itemize}

Summarizing the above analysis we get Eq.~\eqref{eq:C1} in the main text.

\end{document}